\documentclass[12pt,a4paper]{iopart}
\usepackage[english]{babel}

\usepackage{floatrow}
\usepackage{iopams}
\usepackage{verbatim}
\usepackage[usenames,dvipsnames]{xcolor}
\usepackage{graphicx}
\usepackage{float}
\usepackage{color}
\usepackage[colorlinks,citecolor=blue,linkcolor=blue,urlcolor=blue]{hyperref}
\usepackage{amssymb}
\usepackage{hyperref}
\usepackage[utf8]{inputenc}

\def\iotabar{\lower3pt\hbox{$\mathchar'26$}\mkern-7mu\iota}

\newcommand{\rhobf}{\mbox{\boldmath$\rho$}}

\newcommand{\sumsig}{ \raise -1.3ex\hbox{${{\displaystyle \sum}\atop{\scriptstyle \sigma}}$} } 
\newcounter{appnumb}

\newcommand{\gene}{\textsc{Gene}}
\newcommand{\casdk}{{\small CAS3D-K}}

\usepackage{soul}

\begin{document}

\title[Residual zonal flows in tokamaks and stellarators at arbitrary wavelengths]{Residual zonal flows in tokamaks and stellarators at arbitrary wavelengths}
\author{Pedro Monreal$^{1}$} \vspace{-0.2cm}
\eads{\mailto{pedro.monreal@ciemat.es}} \vspace{-0.5cm} 
\author{Iv\'an Calvo$^{1}$} \vspace{-0.2cm}
\eads{\mailto{ivan.calvo@ciemat.es}} \vspace{-0.5cm} 
\author{Edilberto S\'anchez$^{1}$} \vspace{-0.2cm}
\eads{\mailto{edi.sanchez@ciemat.es}} \vspace{-0.5cm} 
\author{F\'elix I Parra$^{2,3}$} \vspace{-0.2cm}
\eads{\mailto{felix.parradiaz@physics.ox.ac.uk}} \vspace{-0.5cm} 
\author{Andr\'es Bustos$^{4,5}$} \vspace{-0.2cm}
\eads{\mailto{anbustos@fis.uc3m.es}} \vspace{-0.5cm} 
\author{Axel K\"onies$^{6}$} \vspace{-0.2cm}
\eads{\mailto{axel.koenies@ipp.mpg.de}} \vspace{-0.5cm} 
\author{Ralf Kleiber$^{6}$} \vspace{-0.2cm}
\eads{\mailto{ralf.kleiber@ipp.mpg.de}} \vspace{-0.5cm} 
\author{Tobias G\"orler$^{4}$} \vspace{-0.2cm}
\eads{\mailto{tbg@ipp.mpg.de}} \vspace{-0.5cm} 

\vspace{0.75cm}

\address{$^1$Laboratorio Nacional de Fusi\'on, CIEMAT, 28040 Madrid, Spain}
\address{$^2$Rudolf Peierls Centre for Theoretical Physics, University of Oxford, Oxford, OX1 3NP, UK}
\address{$^3$Culham Centre for Fusion Energy, Abingdon, OX14 3DB, UK}
\address{$^4$Max-Planck-Institut f\"ur Plasmaphysik, D-85748 Garching, Germany}
\address{$^5$Departamento de F\'isica, Universidad Carlos III de Madrid, 28911 Legan\'es, Spain}
\address{$^6$Max-Planck-Institut f\"ur Plasmaphysik, D-17491 Greifswald, Germany}

\vskip 0.5cm

{\large
\begin{center}
 \today
\end{center}
}

\vspace{-0.5cm}

\begin{abstract}
  In the linear collisionless limit, a zonal potential perturbation in
  a toroidal plasma relaxes, in general, to a non-zero residual
  value. Expressions for the residual value in tokamak and stellarator
  geometries, and for arbitrary wavelengths, are derived.  These
  expressions involve averages over the lowest order particle
  trajectories, that typically cannot be evaluated analytically. In
  this work, an efficient numerical method for the evaluation of such
  expressions is reported. It is shown that this method is faster than
  direct gyrokinetic simulations performed with the
    \gene~and EUTERPE codes.  Calculations of the residual
  value in stellarators are provided for much shorter wavelengths than
  previously available in the literature. Electrons must be treated
  kinetically in stellarators because, unlike in tokamaks, kinetic
  electrons modify the residual value even at long wavelengths. This
  effect, that had already been predicted theoretically, is confirmed
  by gyrokinetic simulations.
\end{abstract}

\maketitle

\section{Introduction}
\label{sec:Introduction}

Whereas the reduction of turbulent transport by zonal flows is a widely accepted phenomenon~\cite{Hammett1993,Diamond2005}, its quantitative understanding is still poor. The determination of the zonal flow amplitude and turbulent transport level are nonlinear questions whose answers require costly gyrokinetic simulations.  The cost grows enormously if, by means of parameter scans, one wants to know how those quantities depend on the magnetic geometry or the plasma conditions.  A useful simplification is provided by the initial value problem consisting of calculating the long-time and collisionless evolution of a zonal perturbation.  Among other reasons, its usefulness is due to the fact that an exact expression for the value of the perturbation at $t=\infty$, called residual value, can be derived.  Even though the explicit evaluation of the  final expression can only be carried out in simplified geometries and for long wavelengths of the perturbation, the analytical solution gives insight into the physics.  
This partly explains the attention attracted by the work of Rosenbluth and Hinton~\cite{Rosenbluth1998} and the effort put on subsequent extensions that we cite below. In stellarator research, the interest in this problem was spurred by the suggestion~\cite{Watanabe2008} of a direct relation between zonal flow residual value and turbulent transport level.

The seminal paper~\cite{Rosenbluth1998} dealt with long-wavelength
potential perturbations in large aspect ratio and circular cross
section tokamaks.  Explicit solutions of related problems for more
complex tokamak geometries and arbitrary wavelengths have been given
in~\cite{Xiao2006,Xiao2007}. In recent years, several articles have
addressed the problem in stellarators~\cite{Sugama2005, Sugama2006,
  Mishchenko2008, Helander2011, Xanthopoulos2011}.  In this paper we
report on a code to evaluate fast and accurately the exact expressions
of the zonal flow residual value, for arbitrary wavelengths and for
tokamak and stellarator geometries.  But why is this useful if the
same answer can be obtained, in principle, by linear runs of a local
gyrokinetic code?  As we will illustrate later on, it turns out that
the solution by means of gyrokinetic simulations of the residual zonal
flow problem, especially at short wavelengths, is very demanding in
terms of computational resources.  We will show that our method is
faster.  These differences can be of several orders of magnitude in
computing time, especially in the case of stellarators. Moreover, the
code and the results of this work are not only interesting from the
point of view of physics, but also for validation of gyrokinetic
codes.

The rest of the paper is organized as follows. In
Section~\ref{sec:Residual} we derive the residual zonal flow
expressions for arbitrary wavelengths and magnetic geometry.  In
Section~\ref{sec:cas3dk} we report on the code used to evaluate the
expressions presented in Section~\ref{sec:Residual}. As a check, we
compare our calculations with analytical results from
\cite{Xiao2006,Xiao2007}, obtained in simplified tokamak geometry. In
order to avoid any confusion, we note that the results in
\cite{Xiao2006,Xiao2007} were not derived as an
  initial value problem, but as the stationary solution of a forced
  system. We explain this in more detail in
Section~\ref{sec:cas3dk}. In Section~\ref{sec:results} our results are
compared with local and global gyrokinetic simulations of the initial
value problem, employing the codes
\gene~\cite{Jenko2000,Gorler2011,GENE,Xanthopoulos2009} and {\small EUTERPE}
\cite{Jost2001, Kleiber2012}, respectively. We also compare the
differences in computational time required by each approach showing
that the method introduced in this paper is faster than
those gyrokinetic simulations. Stellarator residual
values are calculated, using our method and also gyrokinetic codes,
for a range of wavelengths much wider than previously available in the
literature. The stellarator calculations are done for the standard
configuration of Wendelstein 7-X (W7-X). We comment on a purely
stellarator effect already predicted in~\cite{Sugama2006,
  Helander2011}; namely, that the approximation of adiabatic electrons
is always incorrect (even at long wavelengths) for the purpose of
determining the residual zonal flow in stellarators.  We quantify the
error by computing, with our method, the residual value when kinetic
or adiabatic electrons are used. This result is confirmed by
gyrokinetic simulations. The conclusions are presented in
Section~\ref{sec:conclusions}.

Throughout this paper, we assume that the electrostatic potential $\varphi$ associated to the zonal flow is constant on flux surfaces. It is true that zonal flows ({\it i.e.} flows that are only weakly damped, and therefore remain in the plasma for long times) usually correspond to electrostatic potentials with small variations on flux surfaces, but we should emphasize that assuming that $\varphi$ is constant on flux surfaces is, at most, a good approximation.

\section{Linear collisionless evolution of zonal flows}
\label{sec:Residual}

In this section, we give a detailed calculation of the residual zonal flow for arbitrary wavelengths in tokamak and stellarator geometries. We solve the linear and collisionless gyrokinetic equations at long times, assuming that the electrostatic potential perturbation is constant on flux surfaces.

In strongly magnetized plasmas, one employs the smallness of $\rho_{ts\star} =\rho_{ts}/L$ to average over the gyromotion. Here, $L$ is the characteristic length of variation of the magnitude of the magnetic field $B$, $\rho_{ts}=v_{ts}/\Omega_s$ is the thermal gyroradius, $v_{ts} = \sqrt{T_s/m_s}$ is the thermal speed, $\Omega_s=Z_seB/m_s$ is the gyrofrequency, $T_s$ is the equilibrium temperature, $m_s$ is the mass, and $Z_se$ is the charge of species $s$, where $e$ is the proton charge. Gyrokinetic theory~\cite{Catto1978,Frieman1982,Brizard07,Parra2011,Dubin1983,Parra2014} gives a procedure to rigorously derive the gyroaveraged kinetic equations order by order in $\rho_{ts\star} \ll 1$. The averaging operation is conveniently expressed in a new set of phase space coordinates, called gyrokinetic coordinates. Denote by $\{\mathbf{r},\mathbf{v}\}$ the particle position and velocity. The coordinate transformation, to lowest order in $\rho_{ts\star}$, is given by
\begin{eqnarray}\label{eq:gyrokintrans}
  \mathbf{r} = 
  \mathbf{R} + \rhobf_s(\mathbf{R},v,\lambda,\gamma) 
  + O(\rho_{ts\star}^2 L),\nonumber\\[5pt]
  \mathbf{v} = 
  v_{\parallel}(\mathbf{R},v,\lambda,\sigma)\hat\mathbf{b}(\mathbf{R}) + 
  \Omega_s
  \rhobf_s(\mathbf{R},v,\lambda,\gamma)\times\hat\mathbf{b}(\mathbf{R}) 
+ O(\rho_{ts\star}v_{ts}).
\end{eqnarray}
In equation (\ref{eq:gyrokintrans}), $\mathbf{R}$ is the gyrocenter position, $v$ is the magnitude of $\mathbf{v}$, $\lambda = B^{-1}v_\perp^2/v^2$ is the pitch angle, $\sigma = v_{\parallel}/|v_{\parallel}|$ is the sign of the parallel velocity 
\begin{equation}
v_\parallel(\mathbf{R},v,\lambda,\sigma) = 
 \sigma v \sqrt{1- \lambda B(\mathbf{R})},
\end{equation}
$\hat\mathbf{b}$ is the unit vector in the direction of the magnetic field $\mathbf{B}$, $v_\perp$ is the component of the velocity perpendicular to $\mathbf{B}$, and $\rhobf_s$ is the gyroradius vector, defined as
\begin{equation}
\rhobf_s(\mathbf{R},v,\lambda,\gamma) 
 = \frac{m_sv}{Z_s e}\sqrt{\frac{\lambda}{B(\mathbf{R})}}
\left[\hat\mathbf{e}_2(\mathbf{R}) \cos{\gamma} -
\hat\mathbf{e}_1(\mathbf{R}) \sin{\gamma}\right].
\end{equation}
Here, $\hat\mathbf{e}_1(\mathbf{R})$ and $\hat\mathbf{e}_2(\mathbf{R})$ are unit vector fields orthogonal to each other which satisfy $\hat\mathbf{e}_1 \times \hat\mathbf{e}_2 = \hat\mathbf{b}$ at every point. Finally, the gyrophase $\gamma$ is 
\begin{equation}
\gamma = \arctan(\mathbf{v}\cdot \hat\mathbf{e}_2 /
\mathbf{v}\cdot \hat\mathbf{e}_1).
\end{equation}

We introduce straight field line coordinates $\{\psi,\theta,\zeta\}$, where $\psi\in[0,1]$ is the radial coordinate defined as the normalized toroidal flux $\psi=\Psi_t/\Psi_t^\mathrm{edge}$, $\theta$ is a poloidal angle and $\zeta$ is a toroidal angle, with $\theta,\zeta\in[0,1)$. The magnetic field in these coordinates is written as
\begin{equation}\label{eq:Bcontravariantform}
 {\bf B} = 
  -\Psi_p'(\psi)\nabla \psi \times \nabla (\zeta - q(\psi)\theta).
\end{equation}
Here, $q(\psi) = \Psi'_t(\psi)/\Psi'_p(\psi)$ is the safety factor, and $\Psi_p'(\psi)$ and $\Psi_t'(\psi)$ are the derivatives of the poloidal and toroidal fluxes with respect to $\psi$. It will be convenient to define the coordinate $\alpha := \zeta - q(\psi) \theta$, that labels magnetic field lines on each flux surface. Unless otherwise stated, we use $\{\psi,\theta,\alpha\}$ as the set of independent spatial coordinates. Note that $\mathbf{B} \cdot\nabla\psi = 0$, $\mathbf{B} \cdot\nabla\alpha = 0$. We employ $\theta$ as the coordinate along a field line.

The distribution function in gyrokinetic coordinates, $F_s=F_s(\psi,\theta,\alpha,v,\lambda,\sigma,\gamma,t)$, can be written as
\begin{equation}
F_s(\mathbf{R},v,\lambda,\sigma,\gamma,t) = F_{s0}(\mathbf{R},v) +
F_{s1}(\mathbf{R},v,\lambda,\sigma,t) + O(\rho_{ts\star}^2F_{s0}),
\end{equation}
where $F_{s1}=O(\rho_{ts\star}F_{s0})$ and $F_{s0}$ is a Maxwellian distribution whose density $n_s$ and temperature $T_s = m_s v_{ts}^2$ are flux functions,
\begin{equation}
F_{s0}(\mathbf{R},v) := 
 \frac{n_s(\psi(\mathbf{R}))}{(\sqrt{2\pi} v_{ts}(\psi(\mathbf{R})))^3}
 \exp\left(-\, \frac{v^2}{2v^2_{ts}(\psi(\mathbf{R}))}\right).
\end{equation}
The lowest-order quasineutrality condition implies $\sum_s Z_s e n_s(\psi) = 0$. Note that to $O(\rho_{ts\star}F_{s0})$ the distribution function is independent of the gyrophase.

The linear and collisionless time evolution of $F_{s1}$ is given by~\cite{CalvoParra2012,Parra2014}
\begin{eqnarray}\label{eq:gk0}
\fl  \partial_t H_{s1} 
   + (v_{\parallel}\,\hat\mathbf{b} + \mathbf{v}_{ds})
  \cdot\nabla H_{s1}
   = 
\frac{Z_se}{T_s}\partial_t \langle\varphi\rangle F_{s0}
\nonumber \\[5pt]
\fl\hspace{1cm}
+
\frac{1}{B}\left(\nabla\langle\varphi\rangle
\times\hat\mathbf{b}\right)\cdot\nabla\psi
\left[\frac{n_s'}{n_s} 
    + \left(\frac{m_s v^2}{2 T_s} +
      \frac{3}{2}\right)\frac{T_s'}{T_s} \right] F_{s0},
\end{eqnarray}
where primes denote differentiation with respect to $\psi$, the function $H_{s1}$ is defined by
\begin{equation}
H_{s1}= F_{s1} + \frac{Z_se}{T_s}
		\langle\varphi\rangle F_{s0},
\end{equation}
the gyroaveraged electrostatic potential is 
\begin{equation}
 \langle \varphi\rangle\left(\mathbf{R},v,\lambda, t\right):= 
  \frac{1}{2\pi}\int_0^{2\pi} 
  \varphi \left(\mathbf{R}+\rhobf_s(\mathbf{R},v,\lambda,\gamma),t
	  \right)\mathrm{d}\gamma
\end{equation}
and the magnetic drift velocity reads
\begin{equation}
 \mathbf{v}_{ds} =
  \frac{v^2}{\Omega_s} \hat\mathbf{b} \times 
  \left[(1-\lambda B)\hat\mathbf{b}\cdot\nabla
   \hat\mathbf{b} + \frac{\lambda}{2}\nabla B \right].
  \label{eq:driftv}
\end{equation}

The orderings in gyrokinetic theory allow to separate the variations of the fields on the small and large scales, and decompose in Fourier modes with respect to the former. Since we are interested in studying the evolution of an electrostatic potential perturbation that depends only on $\psi$ and the problem is linear, we can take a single mode of the form
\begin{equation}\label{eq:eikonalvarphi}
\varphi(\mathbf{r},t) =
\varphi_k(\psi(\mathbf{r}),t) \exp(\mathrm{i}k_\psi\psi(\mathbf{r})).
\end{equation}
Here,
\begin{equation}
L^{-1}\ll k_\perp \lesssim \rho_{ts}^{-1}
\end{equation}
with $k_\perp(\mathbf{R}) = k_\psi |\nabla \psi(\mathbf{R})|$, and $\varphi_k$ varies on the macroscopic scale $L$. Observe that, due to the effects of magnetic geometry, the dependence of $\varphi_k$ on $\psi$ cannot be avoided even for flat density and temperature profiles. A recent explanation of scale separation, as well as a proof of the equivalence between the local and global approaches to gyrokinetic theory can be found in \cite{Parra2015b}.

To lowest order, the gyroaveraged electrostatic potential is
\begin{equation}
 \langle \varphi \rangle(\mathbf{R},v,\lambda, t)  =
  \varphi_k(\psi(\mathbf{R}),t) \, J_0(k_\perp\rho_s)
\exp({\mathrm{i}k_\psi\psi(\mathbf{R})}),
\end{equation}
where the magnitude of the gyroradius vector is
\begin{equation}
\rho_s(\mathbf{R},v,\lambda) 
= \frac{m_sv}{Z_se}\sqrt{\frac{\lambda}{B(\mathbf{R})}}
\end{equation}
 and $J_0$ is the zeroth-order Bessel function of the first kind,
\begin{equation}
 J_0(x) = 
  \frac{1}{2\pi}\int_0^{2\pi}
  \exp(\mathrm{i} x \sin\gamma)\mathrm{d}\gamma.
\end{equation}
If the electrostatic potential has the form (\ref{eq:eikonalvarphi}), then the distribution function can be written as
\begin{equation}
\fl  F_{s1}(\mathbf{R},v,\lambda,\sigma,t)=
  f_s(\psi(\mathbf{R}),\theta(\mathbf{R}),\alpha(\mathbf{R}),
v,\lambda,\sigma,t) \,\exp({\mathrm{i}k_\psi\psi(\mathbf{R})})
\end{equation}
and consequently
\begin{equation}\label{eq:eikonalH}
\fl H_{s1}(\mathbf{R},v,\lambda,\sigma,t)
= h_s(\psi(\mathbf{R}),\theta(\mathbf{R}),\alpha(\mathbf{R}),
v,\lambda,\sigma,t) \,\exp({\mathrm{i}k_\psi\psi(\mathbf{R})}),
\end{equation}
where $f_s$ and $h_s$ vary on the scale $L$. Then, equation (\ref{eq:gk0}) becomes
\begin{eqnarray}
 \left(\partial_t + v_\parallel\, \hat\mathbf{b} \cdot \nabla 
  + \mathrm{i}k_\psi \omega_s\right) h_s 
 = \frac{Z_s e}{T_s} \partial_t \varphi_k J_{0s} F_{s0},
 \label{eq:gk2}
\end{eqnarray}
where we have used the notation $\omega_s := \mathbf{v}_{ds}\cdot\nabla \psi$ for the radial magnetic drift frequency and $J_{0s}\equiv J_0(k_\perp \rho_s)$. From now on, and for brevity, we omit the dependence of $\varphi_k$ on $\psi$; that is, we write $\varphi_k(t)$ instead of $\varphi_k(\psi,t)$.

Denote by $\omega$ the frequency associated to the time derivative in (\ref{eq:gk2}). The objective is to expand (\ref{eq:gk2}) in powers of $\omega/(v_{ts}L^{-1})\ll 1$, solve the lowest order equations and determine $\varphi_k(t)$ in the limit $t\to \infty$. The $\omega/(v_{ts}L^{-1})\ll 1$ expansion means, in particular, that we average over the lowest order particle trajectories and solve for time scales much longer than a typical orbit time, which is $O(L/v_{ts})$. We define the orbit average for a phase-space function $Q(\psi,\theta,\alpha,v,\lambda,\sigma,t)$ as
\begin{equation}
\overline{Q} :=
 \left\{
  \begin{array}{lcr}
   \langle B\, Q /|v_\parallel |\rangle_\psi /
    \langle B /|v_\parallel|\rangle_\psi &&
    \textrm{for passing particles\,\,}\vspace{0.2cm} \\ 
   \omega_b\oint \mathrm{d}\theta \,Q /(v_\parallel\,
    \hat\mathbf{b} \cdot \nabla \theta) && 
    \textrm{for trapped particles,}
  \end{array}
 \right.
 \label{eq:bounceaverages}
\end{equation}
where $\omega_b := [\oint \mathrm{d}\theta / (v_\parallel\, \hat\mathbf{b}\cdot\nabla \theta)]^{-1}$ is the bounce frequency. Given a function $G(\psi,\theta,\alpha)$, the flux surface average is defined by
\begin{equation}
 \langle G \rangle_\psi = 
  V'(\psi)^{-1} \int_0^{1} \mathrm{d}\theta \int_0^{1} 
  \mathrm{d} \alpha \sqrt{g}\, G(\psi,\theta,\alpha).
  \label{eq:fsa}
\end{equation}
Here, $\sqrt{g} = [(\nabla\psi \times \nabla\theta) \cdot \nabla\alpha]^{-1}$ is the square root of the metric determinant and $V'(\psi) = \int_0^{1} \mathrm{d} \theta \int_0^{1} \mathrm{d} \alpha \sqrt{g}$ is the derivative of the volume enclosed by the flux surface labeled by $\psi$. The symbol $\oint$ stands for integration over the trapped trajectory, where the bounce points $\theta_b$ are the solutions of $1-\lambda B(\psi,\theta_b,\alpha) = 0$ for given values of $\psi$ and $\alpha$, and given an initial condition for the particle trajectory.

Observe that the orbit average operation has the property
\begin{equation}
\overline{v_\parallel \hat\mathbf{b}\cdot\nabla Q} = 0
\end{equation}
for any single-valued function $Q$. We write the radial magnetic drift frequency as a sum of its orbit averaged and fluctuating parts
\begin{equation}
 \omega_s = 
  \overline{\omega_s}
  +v_\parallel\, \hat\mathbf{b}\cdot\nabla \delta_s,
  \label{eq:mde}
\end{equation}
where $\delta_s = \delta_s(\psi,\theta,\alpha,v,\lambda,\sigma)$, that we choose to be odd in $\sigma$, is the radial displacement of the particle's gyrocenter from its mean flux surface. The solution of the magnetic differential equation that determines $\delta_s$ is given in~\ref{sec:mde}.

Defining $\underline{h_s} := h_s \exp({\mathrm{i} k_\psi \delta_s})$ and $\underline{\varphi_k} := \varphi_k \exp({\mathrm{i} k_\psi \delta_s})$, equation (\ref{eq:gk2}) yields
\begin{equation}
 \left( \partial_t  + v_\parallel \hat\mathbf{b}\cdot\nabla + 
  \mathrm{i} k_\psi \overline{\omega_s} \right) 
 \underline{{h}_s} =
 \frac{Z_s e}{T_s} \partial_t\underline{{\varphi}_k} J_{0s} 
 F_{s0}.
 \label{eq:gkUnderlinedVariables}
\end{equation}
It is worth noting that the expansion in $\omega/(v_{ts}L^{-1})$ only makes sense if
\begin{equation}
\frac{k_\psi\overline{\omega_s}}{v_{ts}L^{-1}}\sim
  \frac{\omega}{v_{ts}L^{-1}}
\ll 1.
\end{equation}
For $k_\perp\rho_{ts}\sim 1$, this implies
\begin{equation}\label{eq:condexpansion}
  \frac{\overline{\omega_s}}{\omega_s}
  \sim
  \frac{\omega}{v_{ts}L^{-1}}
  \ll 1.
\end{equation}
This trivially holds in a tokamak because $\overline{\omega_s}=0$ for all trajectories. In a generic stellarator, $\overline{\omega_s}=0$ only for passing particles. Then, condition (\ref{eq:condexpansion}) requires that the secular radial drifts of trapped particles be sufficiently small. We assume that this is the case and carry out the expansion in $\omega/(v_{ts}L^{-1})$.

We write
\begin{equation}
\underline{{h}_s}=\underline{{h}_{s}^{(0)}} +
\underline{{h}_{s}^{(1)}}+
\underline{{h}_{s}^{(2)}}+\dots,
\end{equation}
with $\underline{{h}_{s}^{(j+1)}}/\underline{{h}_{s}^{(j)}}\sim \omega/(v_{ts}L^{-1})$. Then, we expand equation (\ref{eq:gkUnderlinedVariables}). To lowest order, one obtains 
\begin{equation}\label{eq:lowestorderequationh}
 v_\parallel \hat\mathbf{b}\cdot\nabla
  \underline{{h}_{s}^{(0)}} = 0,
\end{equation}
implying that $\underline{{h}_{s}^{(0)}}$ is constant along the lowest order trajectories; {\it i.e.}
\begin{equation}\label{eq:lowestorderequationh}
  \underline{{h}_{s}^{(0)}} = \overline{\underline{{h}_{s}^{(0)}}}.
\end{equation}
To next order, we have
\begin{eqnarray}
 \left( \partial_t + \mathrm{i} k_\psi \overline{\omega_s} \right)
  \underline{{h}_{s}^{(0)}}
  + v_\parallel \hat\mathbf{b}\cdot\nabla
  \underline{{h}_{s}^{(1)}} = 
  \frac{Z_s e}{T_s} \partial_t \underline{{\varphi}_k}
  J_{0s} F_{s0}. 
  \label{eq:nextordereqNotInLaplaceSpace}
\end{eqnarray}
We do not write $\underline{{\varphi}_{k}^{(0)}}$ to ease the notation. The orbit average of (\ref{eq:nextordereqNotInLaplaceSpace}) annihilates the term $v_\parallel \hat\mathbf{b}\cdot\nabla \underline{{h}_{s}^{(1)}}$, and we find
\begin{eqnarray}
 \left( \partial_t + \mathrm{i} k_\psi \overline{\omega_s} \right)
  \underline{{h}_{s}^{(0)}}
   = 
  \frac{Z_s e}{T_s} \overline{\partial_t \underline{{\varphi}_k}
  J_{0s}} F_{s0}. 
  \label{eq:nextordereqNotInLaplaceSpaceAveraged}
\end{eqnarray}
It is useful to work in Laplace space in order to solve this equation. The Laplace transform of a function $Q(t)$ is defined as $\widehat{Q}(p) = \int_0^\infty Q(t) \mathrm{e}^{-pt} \mathrm{d}t$, where $p$ denotes the variable in Laplace space. We apply it to (\ref{eq:nextordereqNotInLaplaceSpaceAveraged}) and obtain
\begin{eqnarray}
 \left( p + \mathrm{i} k_\psi \overline{\omega_s} \right)
  \underline{\widehat{h}_{s}^{(0)}}
   = 
  \frac{Z_s e}{T_s}p \overline{\underline{\widehat{\varphi}_k}
  J_{0s}} F_{s0} + \underline{f_s}(0). 
  \label{eq:nextordereqInLaplaceSpaceAveraged}
\end{eqnarray}
Here, $\underline{f_s}(0) := f_s(0) \exp({ \mathrm{i} k_\psi \delta_s})$ and $f_s(0)$ is the initial condition for $f_s$; {\it i.e.} $f_s(0)\equiv f_s(\psi,\theta,\alpha,v,\lambda,\sigma,0)$.

The solution of (\ref{eq:nextordereqInLaplaceSpaceAveraged}) yields
\begin{equation}
\widehat{h}_s^{(0)} = 
  \frac{\mathrm{e}^{-\mathrm{i} k_\psi \delta_s}}{p + \mathrm{i} k_\psi
  \overline{\omega_s}} 
  \left( \frac{Z_s e}{T_s} p\, \widehat{\varphi}_k\,
   \overline{\mathrm{e}^{\mathrm{i} k_\psi \delta_s} J_{0s}} F_{s0} 
   + \overline{\mathrm{e}^{\mathrm{i} k_\psi \delta_s} f_s(0) }
  \right).
  \label{eq:gk6}
\end{equation}

In order to have a closed system of equations we employ the gyrokinetic quasineutrality equation (see, for example, references \cite{CalvoParra2012, Parra2014}),
\begin{eqnarray}
\fl  \sum_s \frac{Z_s^2 e}{T_s} n_s\, \varphi(\mathbf{R},t) = \nonumber\\
\fl\hspace{1cm}
  \sum_s Z_s \int 
  H_{s}(\mathbf{R}-\rhobf_s(\mathbf{R},v,\lambda,\gamma),v,\lambda,\sigma,t) 
\mathrm{d}^3 v.
  \label{eq:qn0}
\end{eqnarray}
Here, the short-hand notation $\int Q \mathrm{d}^3 v$ means, for a function $Q(\psi,\theta,\alpha,v,\lambda,\sigma,\gamma)$, 
\begin{equation} 
\fl \int Q \, \mathrm{d}^3\upsilon = 
  \sum_{\sigma=-1}^1
\int_0^{2\pi}\mathrm{d}\gamma\int_0^\infty \mathrm{d}v\, \int_0^{1/B}
  \mathrm{d}\lambda\, \frac{ v^2 B}{2\sqrt{1-\lambda B}}
  Q(\psi,\theta,\alpha,v,\lambda,\sigma,\gamma).
\end{equation}
Using (\ref{eq:eikonalvarphi}) and (\ref{eq:eikonalH}), and flux-surface averaging, we get
\begin{equation}
 \sum_s \frac{Z_s^2 e}{T_s}n_s\, {\varphi}_k = 
  \left\langle 
   \sum_s Z_s \int J_{0s} {h}_s
    \mathrm{d}^3 v  
  \right\rangle_\psi.
  \label{eq:qn2}
\end{equation}
To lowest order in $\omega/(v_{ts}L^{-1})\ll 1$, and after transforming to Laplace space, equation (\ref{eq:qn2}) gives
\begin{equation}
 \sum_s \frac{Z_s^2 e}{T_s}n_s\, \widehat{\varphi}_k = 
  \left\langle 
   \sum_s Z_s \int J_{0s}  \widehat{h}_s^{(0)}
    \mathrm{d}^3 v  
  \right\rangle_\psi.
  \label{eq:qn2aux}
\end{equation}
We employ (\ref{eq:gk6}) to write the right side of (\ref{eq:qn2aux}) in terms of the electrostatic potential and the initial condition, and solve for $\widehat{\varphi}_k$. The result is
\begin{eqnarray}
\fl
\widehat{\varphi}_k(p) = 
 \frac{
 \sum_s Z_s\left\{ \frac{1}{p+\mathrm{i}k_\psi\overline{\omega_s}}
	    \mathrm{e}^{-\mathrm{i}k_\psi\delta_s} J_{0s} 
	    \overline{\mathrm{e}^{\mathrm{i}k_\psi\delta_s} f_s(0) } /F_{s0}
	  \right\}_s
 }{
 \sum_s \frac{Z_s^2 e}{T_s} 
 \left\{1 - \frac{p}{p+\mathrm{i}k_\psi\overline{\omega_s}} \,
  \mathrm{e}^{-\mathrm{i}k_\psi\delta_s} J_{0s} 
  \overline{ \mathrm{e}^{\mathrm{i}k_\psi\delta_s} J_{0s}} \right\}_s  
 },
 \label{eq:qn3}
\end{eqnarray}
where we have simplified the notation by defining
\begin{equation}
\fl
\left\{Q\right\}_s  := 
 \left\langle 
  \sum_{\sigma=-1}^1 \int_0^\infty \mathrm{d}v\,\int_0^{1/B} 
  \mathrm{d}\lambda\, \frac{\pi v^2 B}{\sqrt{1-\lambda B}}
  Q (\psi,\theta,\alpha,v,\lambda,\sigma)\,
  F_{s0} \right\rangle_\psi
 \label{eq:corchete}
\end{equation}
for gyrophase independent functions on phase space.

The residual value is found from the well-known property of the Laplace transform 
\begin{equation}\label{eq:propertyLaplaceTransform}
  \lim_{t\to\infty}\varphi_k(t) = \lim_{p\to 0}p\widehat{\varphi}_k(p).
\end{equation}
Applying (\ref{eq:propertyLaplaceTransform}) to equation (\ref{eq:qn3}), we find
\begin{equation}
 \varphi_k (\infty) = 
  \frac{ \sum_s Z_s\left\{ \mathrm{e}^{-\mathrm{i}k_\psi\delta_s} J_{0s} 
   \overline{\mathrm{e}^{\mathrm{i}k_\psi\delta_s} f_s(0) } /F_{s0}
		   \right\}_s^{\overline{\omega_s}=0}   
  }{
  \sum_s \frac{Z_s^2e}{T_s} 
  \left[\left\{1\right\}_s-\left\{ \mathrm{e}^{-\mathrm{i}k_\psi\delta_s} J_{0s}
	 \overline{ \mathrm{e}^{\mathrm{i}k_\psi\delta_s} J_{0s}}
	 \right\}_s^{\overline{\omega_s}=0} 
  \right]
  },
  \label{eq:rsk}
\end{equation}
where $\varphi_k(\infty) \equiv \lim_{t\to\infty}\varphi_k(t)$. The superscript $\overline{\omega_s}=0$ means that the integration is performed only for particles whose trajectory satisfies $\overline{\omega_s}=0$.  In tokamaks, this property holds true for both trapped and passing particles, and therefore the integrals in (\ref{eq:rsk}) are performed over the whole phase space. In stellarators, $\overline{\omega_s}=0$ is satisfied exclusively for passing particles. Only in perfectly omnigenous stellarators~\cite{Cary1997.1,Cary1997.2,Parra2015} have trapped particles vanishing average radial magnetic drift. Hence, in a generic stellarator, the integrals in (\ref{eq:rsk}) with superscript $\overline{\omega_s}=0$ are performed only over the passing region of phase space.

The residual level is usually defined as the normalized value $\varphi_k(\infty)/\varphi_k(0)$. The relation between $f_s(0)$ and $\varphi_k(0)$ is given by the flux-surface averaged quasineutrality equation at $t=0$,
\begin{equation}
\fl
\sum_s \frac{Z_s^2 e}{T_s}n_s 
  \left\langle 1-\Gamma_0(k_\perp^2\rho_{ts}^2)\right\rangle_\psi \varphi_k(0) =
  \left\langle \sum_s Z_s \int  J_{0s}f_s(0)  \mathrm{d}^3 v \right\rangle_\psi.
   \label{eq:qn4}
\end{equation}
Here, we have employed the identity $\int J_0^2(k_\perp \rho_s) F_{0s} \mathrm{d}^3 v = n_s \Gamma_0(k_\perp^2\rho_{ts}^2)$, where $\Gamma_0(k_\perp^2\rho_{ts}^2) := \mathrm{e}^{-k_\perp^2\rho_{ts}^2} \,I_0(k_\perp^2\rho_{ts}^2)$, and $I_0$ is the zeroth order modified Bessel function. We will use the notation $\Gamma_{0s}\equiv \Gamma_0(k_\perp^2\rho_{ts}^2)$.

For later comparison with gyrokinetic simulations, it will be useful to have at hand the expressions corresponding to the approximation of adiabatic electrons. Using this approximation, equation (\ref{eq:qn2}) can be written as
\begin{equation}
 \sum_{s\neq e}\frac{Z_s^2 e}{T_s} n_s\, \widehat{\varphi}_k(p) = 
  \left\langle \sum_{s\neq e} Z_s\int \mathrm{d}^3 \upsilon\, 
   J_{0s}\widehat{h}_s (p) \right\rangle_\psi.
\end{equation}
Proceeding as shown previously for the fully kinetic case, we find that the expression for $\varphi_k(\infty)$ reads 
\begin{equation}
 \varphi_k (\infty) =
  \frac{ \sum_{s\neq e} Z_s \left\{ \mathrm{e}^{-\mathrm{i}
			     k_\psi\delta_s} J_{0s}  
	    \overline{ \mathrm{e}^{\mathrm{i}k_\psi\delta_s}
	    f_s(0)}/F_{s0} \right\}_s^{\overline{\omega_s}=0}  
  }{
  \sum_{s\neq e} \frac{Z_s^2e}{T_s} 
  \left[\left\{1\right\}_s-\left\{ \mathrm{e}^{-\mathrm{i} k_\psi
	 \delta_s} J_{0s} 
	 \overline{ \mathrm{e}^{\mathrm{i}k_\psi\delta_s} J_{0s}}
			   \right\}_s^{\overline{\omega_s}=0} \right]
  } .
  \label{eq:rsa}
\end{equation}
In this case, the equation that relates $f_s(0)$ and  $\varphi_k(0)$ is 
\begin{equation}
\fl \sum_{s\neq e}\frac{Z_s^2e}{T_s}\, n_s 
  \left\langle 1-\Gamma_{0s}\right\rangle_\psi \varphi_k(0) =
  \left\langle \sum_{s\neq e} Z_s \int 
   J_{0s} f_s(0)  \mathrm{d}^3 v \right\rangle_\psi.
  \label{eq:qn6}
\end{equation}

The residual zonal flow was computed in reference \cite{Rosenbluth1998} for $k_\perp \rho_{ti}\ll 1$ in a large aspect ratio, circular cross section tokamak with adiabatic electrons. An extension of the derivation of \cite{Rosenbluth1998} was proposed in references \cite{Xiao2006, Xiao2007} to allow for short-wavelength perturbations, and also for kinetic electrons and more complex tokamak geometries. In reference \cite{Xiao2007}, comparisons of analytical calculations with gyrokinetic simulations are shown. The enhancement of tokamak residual zonal flows at short wavelengths was originally found in reference \cite{Jenko2000} by means of gyrokinetic simulations with the code {\small GS2}. But the short-wavelength calculations of \cite{Jenko2000, Xiao2006, Xiao2007} do not correspond to an initial zonal value problem because the quasineutrality equation is forced with a source term. At long wavelengths, the initial value problem and the forced system give the same result. This will be explained in more detail 
in Section \ref{sec:cas3dk}.

In stellarators, the residual zonal flow calculation has been done in \cite{Sugama2005,Sugama2006,Mishchenko2008, Helander2011,Xanthopoulos2011}. In references  \cite{Mishchenko2008,Helander2011,Xanthopoulos2011} the emphasis is put on long-wavelength zonal flows. In references \cite{Sugama2005,Sugama2006}, the derivation of the equations is valid for long and short-wavelengths but some approximations are used to describe the magnetic geometry.

\section{Evaluation of the expressions for the residual zonal flow}
\label{sec:cas3dk}

The evaluation of the right side of equation (\ref{eq:rsk}) (and, of course, (\ref{eq:rsa})) requires the calculation of quantities of the form
\begin{equation}
 \left\{P\, \overline{Q}\right\}_s := 
  \left\langle \sum_{\sigma=-1}^1 \int_0^\infty \mathrm{d}v\,
   \int_0^{1/B} \mathrm{d}\lambda\, 
   \frac{\pi v^2 B}{\sqrt{1-\lambda B}}\, P\, \overline{Q}\,
   F_{s0} \right\rangle_\psi,
  \label{eq:corchete2}
\end{equation}
for functions $P=P(\psi,\theta,\alpha,v,\lambda,\sigma)$ and $Q=Q(\psi,\theta,\alpha,v,\lambda,\sigma)$, where the orbit average is defined in (\ref{eq:bounceaverages}).  These averages depend on the details of the magnetic field and cannot be evaluated analytically, except in simplified cases (for example, in the cases considered in \cite{Rosenbluth1998,Xiao2006,Xiao2007}). In this work, we evaluate equation (\ref{eq:rsk}) using the framework of the code {\casdk} \cite{koenies2000,koenies2008}.  For this purpose, we have included in this code the relevant finite Larmor radius effects, the solution to the magnetic differential equations described in \ref{sec:mde}, and the integration over the velocity coordinates $v$ and $\sigma$.

The code {\casdk} is well suited to perform the average (\ref{eq:bounceaverages}).  Since the lowest order particle trajectories lie entirely on flux surfaces, all the calculations are local, thus permitting a parallelization by flux surface using MPI. The magnetic equilibrium is obtained from the 3D MHD equilibrium code {\small VMEC}~\cite{Hirshman1983} and then transformed into Boozer coordinates $\{\psi,\theta,\zeta\}$, which can easily be transformed to the coordinates $\left\{ \psi, \theta, \alpha \right\}$. On a given flux surface, the pitch angle $\lambda$ distinguishes between passing and trapped trajectories.  The passing-trapped boundary is given by $\lambda_\mathrm{\,c} = 1/B^\mathrm{\,max}$, where $B^\mathrm{\,max}$ is the maximum of $B$ on the flux surface. Passing particles have $\lambda$ values with $0 < \lambda < \lambda_\mathrm{c}$ and trapped particles are those with $\lambda_\mathrm{c} < \lambda < 1/B^\mathrm{min}$, where $B^\mathrm{min}$ is the minimum of $B$ on the flux surface.  Trapped 
particles can live inside one or several magnetic field periods. In {\casdk}, they are grouped by the number of periods they go through. The groups are obtained by setting a large number of initial conditions for the trajectories, and finding the bounce points $\theta_b$ from the bounce condition $1-\lambda B(\psi, \theta_b, \alpha) = 0$ for constant $\psi$ and $\alpha$. From this procedure, the boundaries of each group are found and the numerical integration for a given group is performed by covering the region they define with new trajectories.  Note that each group requires different numerical resolution.  Some trapped trajectories close to the passing-trapped boundary may require a large number of periods until the bounce points are found. If this number is sufficiently large, typically larger than 500 periods, the trajectories are then considered as passing. A Gauss-Legendre quadrature scheme is used for the integration in $\theta$, $\alpha$ and $\lambda$, which avoids the numerical problems that may 
appear at points where $1-\lambda B=0$.

In reference \cite{Mishchenko2008}, the phase space
  integrations with {\casdk} discussed in the previous paragraph were
  employed to obtain the zonal flow frequency in stellarator geometry
  in the long-wavelength limit.  In this limit, there are some
  simplifications. We have implemented new functionalities in {\casdk}
  that also allow to deal  with short wavelengths. Next, we describe
  the main features of this extension of {\casdk}.

First, we note that the integration of the resulting
  expressions over the velocity coordinate $v$ could be performed
  analytically in \cite{Mishchenko2008}. If one wants to calculate the
  averages involved in (\ref{eq:rsk}), the integration over the
  velocity coordinates $v$ and $\sigma$ must be computed numerically.
  We have included in {\casdk} the integration over these coordinates on
  top of the $\theta$, $\alpha$ and $\lambda$ integrations. For the
  integration over $v$, a linear scheme has been used. This scheme
  allows the computation of any moment in $v$ of the
  Maxwellian distribution function, $F_{s0}(\psi,v)$.

Second, the equations to obtain the residual level
  (\ref{eq:rsk}) and (\ref{eq:rsa}) incorporate finite Larmor
  radius effects. These effects are encoded in $J_0(k_\perp\rho_s)$,
  $\Gamma_{0} (k_\perp^2\rho_{ts}^2)$ and $\exp(\pm
  \mathrm{i}k_\psi\delta_s)$. Here, $k_\perp=k_\psi |\nabla \psi|$
  where $k_\psi$ is an input parameter and the quantity $|\nabla
  \psi|=|\nabla \psi|(\theta,\alpha)$ is obtained from the {\small
    VMEC} equilibrium. We have included those factors and adapted the
  resolution of the integration on phase space to their strong
  oscillatory behavior at short wavelengths. We have also implemented
  in the code the expression of $\delta_s$ in stellarator
  geometry, for both passing and trapped particles (see the
  derivations in \ref{sec:mde}). For passing particles, $\delta_s$ is
  given by equation (\ref{eq:deltap}) and for trapped particles it is
  given by equation (\ref{eq:mdet3}).

The above modifications included in {\casdk} allow us
  to calculate the residual level in tokamak or stellarator geometry
  for arbitrary wavelengths. In the rest of the paper, we use the
  terminology ``{\casdk}'' and ``extension to {\casdk}'' interchangeably.

\begin{figure}
\includegraphics[width=1.\linewidth]{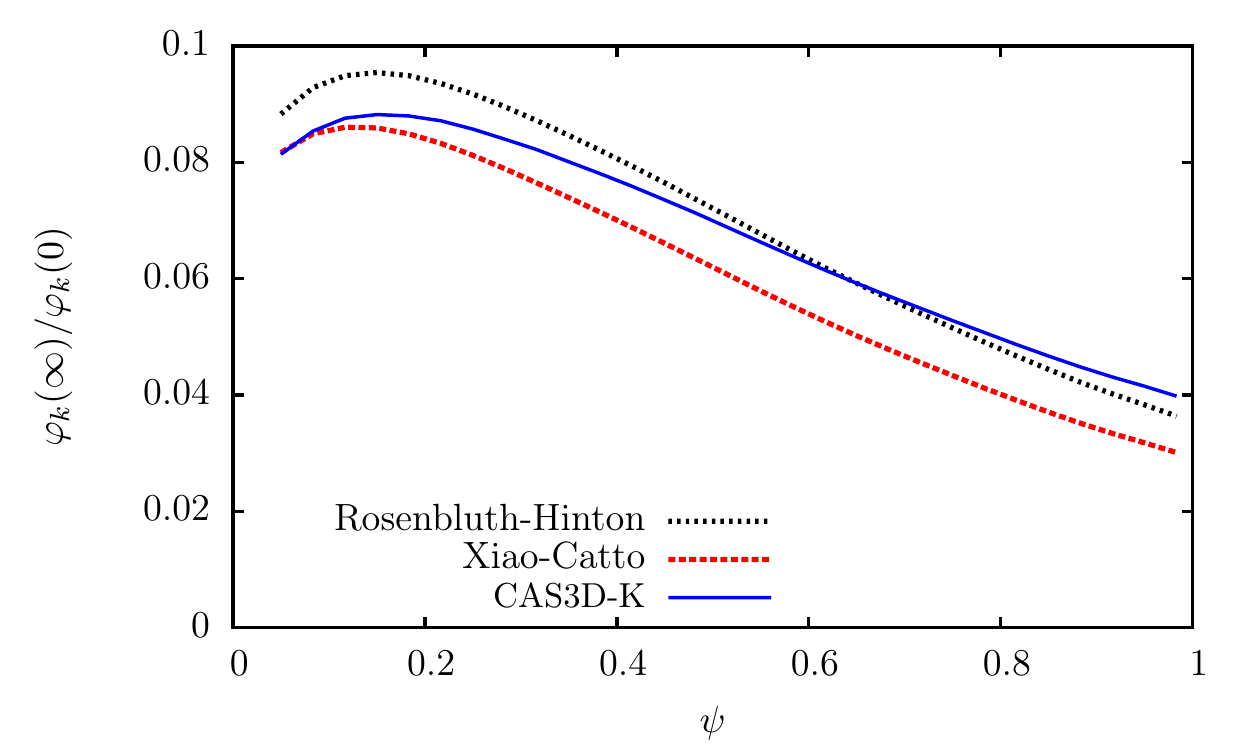}
\caption{Radial dependence of the residual level given by (\ref{eq:rh}), by (\ref{eq:XCkequaltozero}), and by the evaluation of (\ref{eq:rsa}) with {\footnotesize CAS3D-K} in the long-wavelength limit. A tokamak with major radius $R=1.7$~m, minor radius $a=0.4$~m, and $q$ profile given in figure \ref{fig:qprofileSectionAnalyticalChecks} has been used.}
\label{fig:RHXC}
\end{figure}

As a preliminary check of these extensions to {\casdk}, in this section we compare our results with analytical results available in the literature. For these comparisons, we take a plasma consisting of singly charged ions and electrons, and assume flat density and temperature profiles with the same values for both species.

\begin{figure}
\includegraphics[width=1.\linewidth]{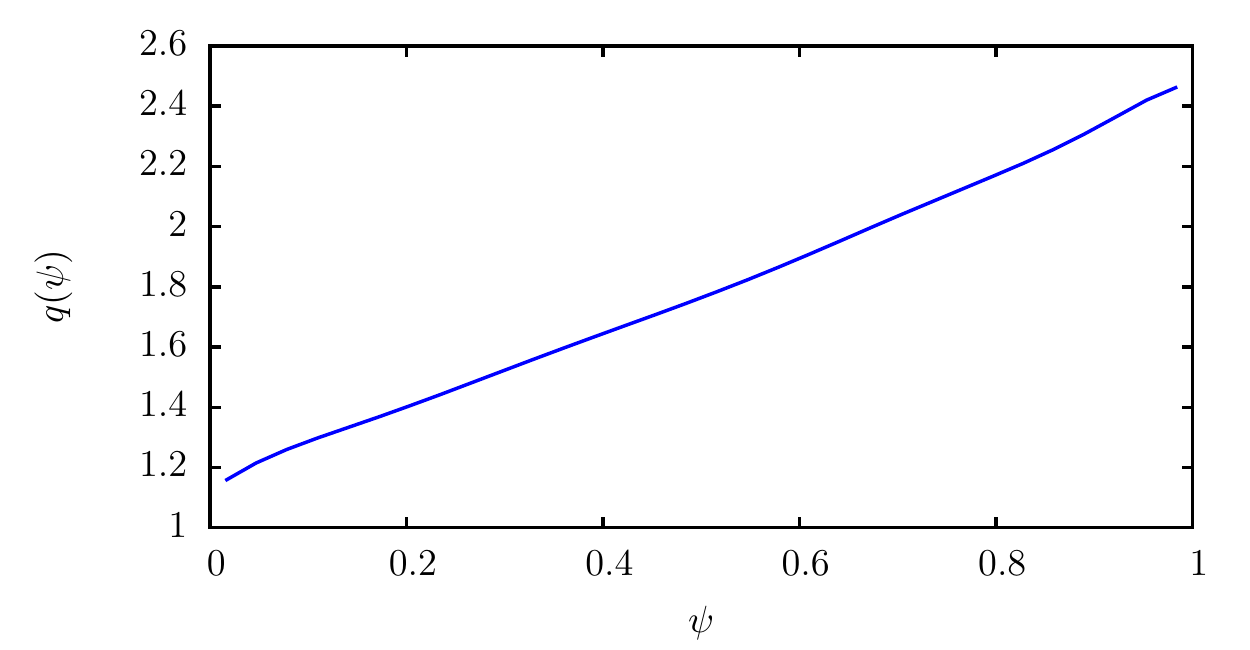}
\caption{Safety factor profile of the tokamak employed for the calculations of Section \ref{sec:cas3dk}.}
\label{fig:qprofileSectionAnalyticalChecks}
\end{figure}

In reference \cite{Rosenbluth1998}, Rosenbluth and Hinton (R-H) calculated the residual level in large aspect ratio tokamaks with circular cross section and adiabatic electrons, in the limit $k_\perp\rho_{ti}\ll 1$. Denote the safety factor by $q$ and the inverse aspect ratio by $\varepsilon = (a/R) \sqrt{\psi}$, where $a$ is the minor radius and $R$ is the major radius. The result obtained in \cite{Rosenbluth1998} is
\begin{equation}
\frac{\varphi_k(\infty)}{\varphi_k(0)} =
  \frac{1}{1+1.6\, q^2\varepsilon^{-1/2}}\,.
  \label{eq:rh}
\end{equation}
In reference \cite{Xiao2006}, Xiao and Catto gave an expression more accurate in the inverse aspect ratio expansion. Namely,
\begin{equation}\label{eq:XCkequaltozero}
  \frac{\varphi_k(\infty)}{\varphi_k(0)}= 
  \frac{1}{1 + 1.6 q^2\varepsilon^{-1/2}+0.5
    q^2 + 0.36 q^2 \varepsilon^{1/2}}\, .
\end{equation}
Both of these results were obtained
by using the analytical equilibrium of a large
  aspect ratio circular tokamak which, in our coordinates, is given by
\begin{equation}
B=\frac{B_0}{1+\varepsilon\cos(2\pi\theta)},
\label{eq:larteq}
\end{equation}
where $B_0$ is the magnetic field strength at the magnetic axis. The
analytical solutions (\ref{eq:rh}) and (\ref{eq:XCkequaltozero}) are
plotted in figure \ref{fig:RHXC}, together with the numerical
evaluation of (\ref{eq:rsa}) with {\casdk}, for $k_\perp\rho_{ti}\ll 1$
and a Maxwellian initial condition. We use an axisymmetric tokamak
with major radius $R=1.7$~m, minor radius $a=0.4$~m, and $q$ profile
given in figure \ref{fig:qprofileSectionAnalyticalChecks}.  For the
{\casdk} computations, the equilibrium is obtained with {\small VMEC}
employing the aspect ratio and safety factor values just mentioned.
 The wavenumber used in the {\casdk} calculation is
  $k_\psi=0.5$ and the dimensionless quantity $\left\langle
    k_\perp\rho_{ti} \right\rangle_\psi$ ranges from 0.0015 in the
  innermost radial position to 0.0068 in the outermost
  one. We have checked that the residual zonal flow
    value obtained with {\casdk} and shown in figure \ref{fig:RHXC} does
    not change if $\left\langle k_\perp\rho_{ti} \right\rangle_\psi$
    is further decreased.  The regions of figure \ref{fig:RHXC}
where the curves agree and where the curves differ are as expected
(see the remarks in \cite{Yamagishi2012} about figure 3(a) in that
reference). The analytical equilibrium of a large aspect ratio
circular tokamak, used in deriving the equations (\ref{eq:rh}) and
(\ref{eq:XCkequaltozero}), differs less from the numerical equilibrium
obtained with {\small VMEC} in radial positions closer to the center.
We will see in Section \ref{sec:results} that the {\casdk} results
coincide with gyrokinetic simulations of zonal flow evolution, in
which {\small VMEC} equilibria are also used.

\begin{figure}
\includegraphics[width=1.\linewidth]{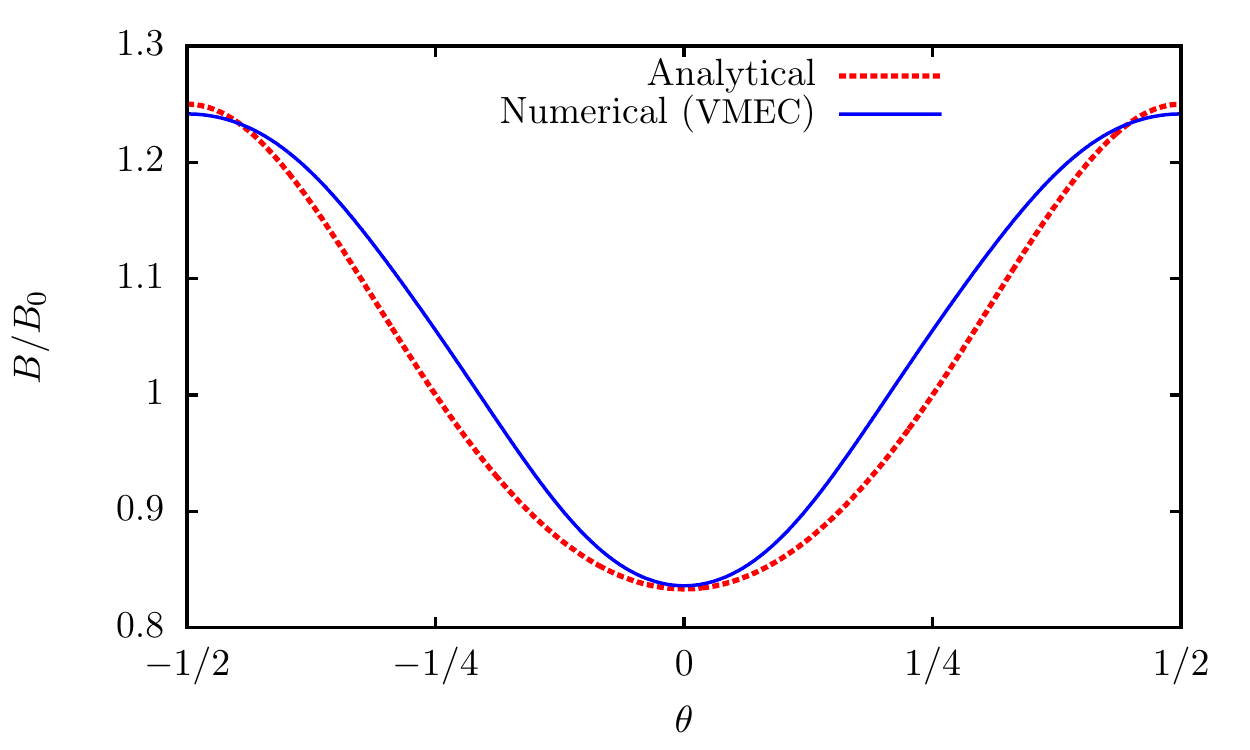}
\caption{Magnetic field strength along a field line of the analytical large aspect ratio circular tokamak equilibrium, given by equation (\ref{eq:larteq}), with $\varepsilon=0.2$, and for the numerical equilibrium obtained with {\footnotesize VMEC} (with $R=1.7$ m, $a=0.4$ m) at $\psi=0.7$.}
\label{fig:bfield}
\end{figure}

As explained above, Xiao and Catto (X-C) also addressed in references \cite{Xiao2006, Xiao2007} the extension of the calculation in \cite{Rosenbluth1998} to short wavelengths. They gave the result
\begin{equation}
 \frac{\varphi_k(\infty)}{\varphi_k(0)} =
  \frac{\sum_s\frac{Z_s^2}{T_s} \left\{1- J_{0s}^2\right\}_s
  }{
  \sum_s \frac{Z_s^2}{T_s} 
  \left\{1 - \mathrm{e}^{-\mathrm{i}k_\psi\delta_s} J_{0s}
	\overline{ \mathrm{e}^{\mathrm{i}k_\psi\delta_s} J_{0s}
			    }\right\}_s}
  \label{eq:xck}
\end{equation}
for the residual zonal flow in a tokamak at arbitrary wavelengths and with kinetic electrons. The expression provided by X-C for the case of adiabatic electrons is
\begin{equation}
\frac{\varphi_k(\infty)}{\varphi_k(0)} =
 \frac{\sum_{s\neq e}\frac{Z_s^2}{T_s} \left\{1- J_{0s}^2\right\}_s
 }{
 \sum_{s\neq e} \frac{Z_s^2}{T_s} 
 \left\{1 - \mathrm{e}^{-\mathrm{i}k_\psi\delta_s} J_{0s} 
   \overline{ \mathrm{e}^{\mathrm{i}k_\psi\delta_s} J_{0s} }\right\}_s}.
 \label{eq:xca}
\end{equation}

\begin{figure}
\includegraphics[width=1.\linewidth]{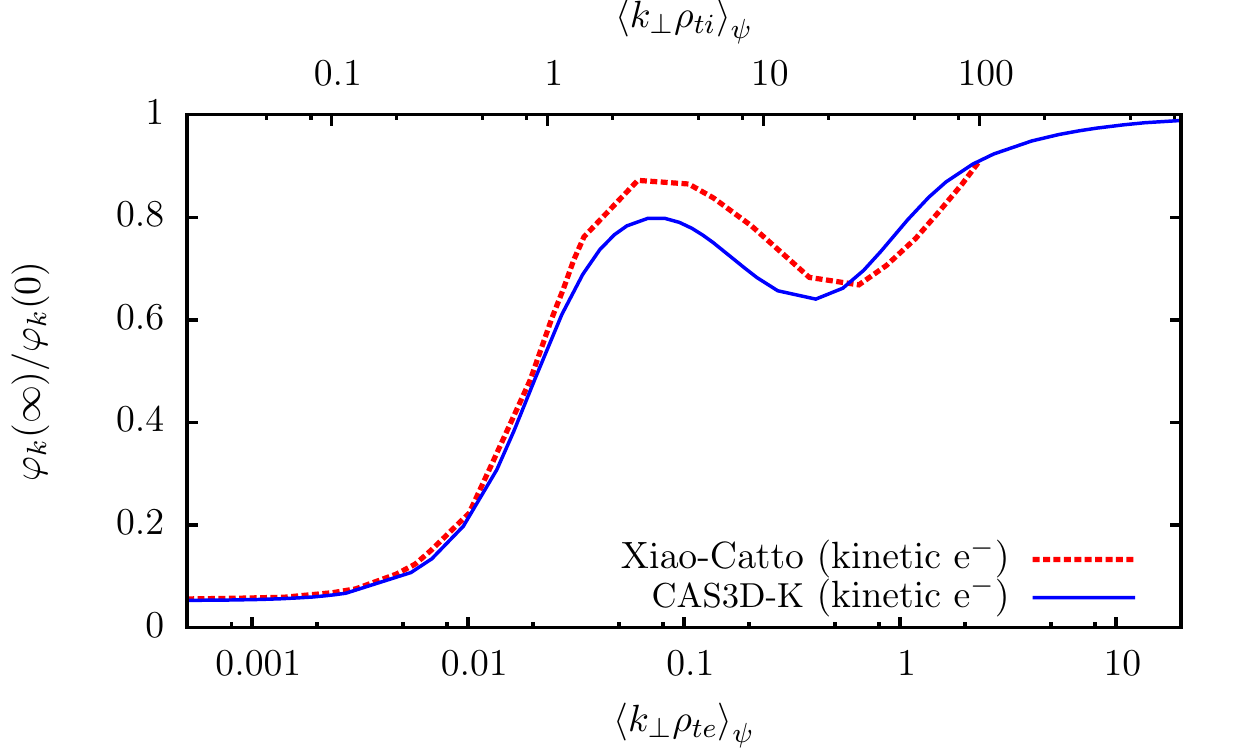}
\caption{Comparison of the result in references \cite{Xiao2006,Xiao2007} and the evaluation of (\ref{eq:xck}) with {\footnotesize CAS3D-K}. The parameters of the tokamak are the same as in figure \ref{fig:RHXC}.}
\label{fig:XC2}
\end{figure}

In order to avoid any confusion, we have to point out that
(\ref{eq:xck}) and (\ref{eq:xca}) were not derived as
  the solution of the initial value problem explained in Section
  \ref{sec:Residual}, but assuming that the
  quasineutrality equation is forced with a source term. The argument
  of X-C can be streamlined as follows. Go back to (\ref{eq:rsk}) for
  the tokamak case (that is, $\overline{\omega_s}=0$ for all
  particles). X-C consider that finite orbit width effects do not
  affect the initial condition; {\it i.e.}
\begin{eqnarray}
\fl
\sum_s Z_s\left\{ \mathrm{e}^{-\mathrm{i}k_\psi\delta_s} J_{0s} 
    \overline{ \mathrm{e}^{\mathrm{i}k_\psi\delta_s} f_s(0)/F_{s0}} \right\}_s
\approx
\nonumber\\[5pt]
\fl
\hspace{1cm}
\sum_s Z_s\left\{J_{0s} f_s(0)/F_{s0}  \right\}_s =
\left\langle \sum_s Z_s \int  J_{0s} f_s(0)  \mathrm{d}^3 v \right\rangle_\psi .
  \label{eq:XCapprox}
\end{eqnarray}
Hence, in this approximation, (\ref{eq:rsk}) gives
\begin{equation}
 \varphi_k (\infty) \approx
  \frac{\left\langle \sum_s Z_s \int  J_{0s} f_s(0)
  \mathrm{d}^3 v \right\rangle_\psi
  }{
  \sum_s \frac{Z_s^2e}{T_s} 
  \left[\left\{1\right\}_s-\left\{ \mathrm{e}^{-\mathrm{i} k_\psi \delta_s} J_{0s} 
	 \overline{ \mathrm{e}^{\mathrm{i} k_\psi \delta_s} J_{0s}}
	   \right\}_s^{\overline{\omega_s}=0} \right]
  }.
  \label{eq:XCapprox}
\end{equation}
The quasineutrality equation at $t=0$, (\ref{eq:qn4}), can be trivially rewritten as
\begin{equation}
\fl
\varphi_k(0) =
\left(\sum_s \frac{Z_s^2 e}{T_s}\{1-J_{0s}^2\}_s\right)^{-1}
  \left\langle \sum_s Z_s \int  J_{0s} f_s(0)  \mathrm{d}^3 v 
\right\rangle_\psi.
   \label{eq:qnt0again}
\end{equation}
From the quotient of (\ref{eq:XCapprox}) and (\ref{eq:qnt0again}), one
obtains equation (\ref{eq:xck}). Analogously, one can obtain
(\ref{eq:xca}) from (\ref{eq:rsa}). From these manipulations, it is
clear that in the X-C calculation the charge perturbation at $t=0$ can
be viewed as a constant source term in the
  quasineutrality equation (see also the remark after equation
  (\ref{eq:rlta})).

References \cite{Xiao2006} and \cite{Xiao2007}
  provided analytical evaluations of the right sides of (\ref{eq:xck})
  and (\ref{eq:xca}) for simplified tokamak geometries. Since we can
  directly evaluate the right sides of (\ref{eq:xck}) and
  (\ref{eq:xca}) with {\casdk}, we will compare the results as an
  additional check of our numerical tool.

\begin{figure}
\includegraphics[width=1.\linewidth]{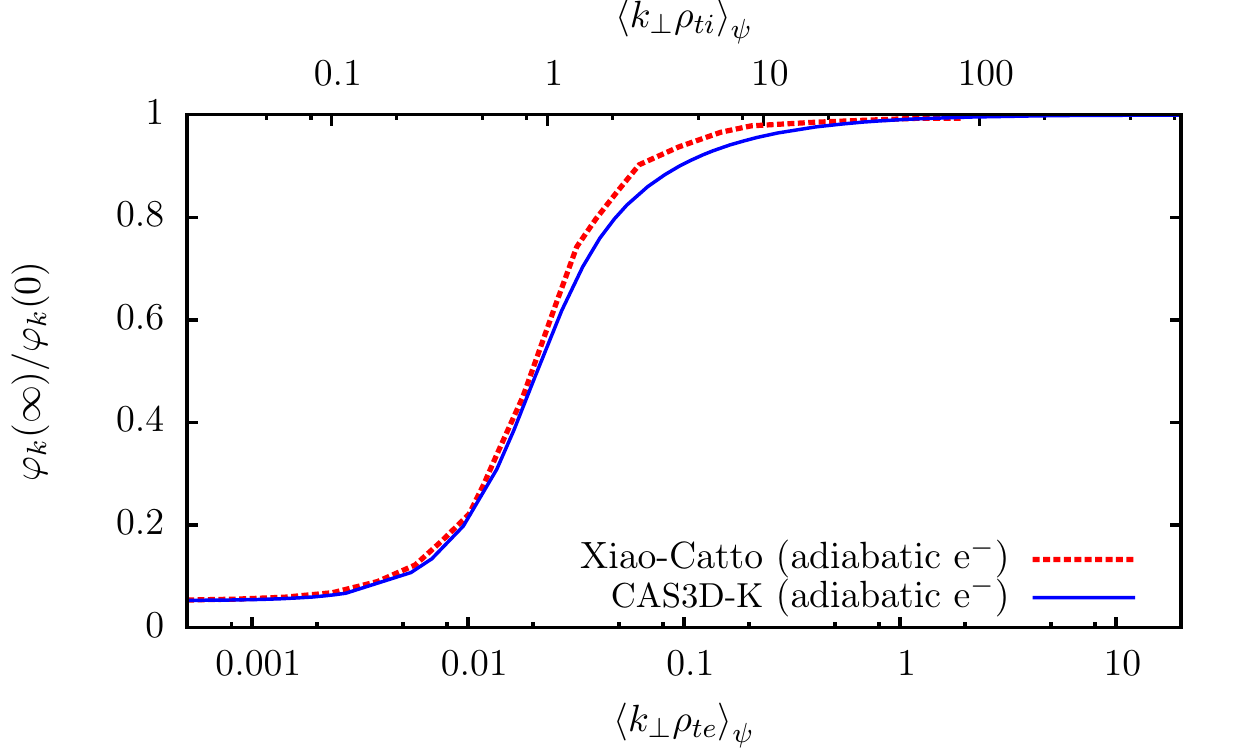}
\caption{Comparison of the result in reference \cite{Xiao2006} for adiabatic electrons and the evaluation of equation (\ref{eq:xca}) with {\footnotesize CAS3D-K}. The parameters of the tokamak are the same as in figure \ref{fig:RHXC}.}
\label{fig:XC1}
\end{figure}

In \cite{Xiao2006,Xiao2007}, the analytical equilibrium of a circular
cross section, large aspect ratio tokamak with safety factor $q=2$ and
inverse aspect ratio $\varepsilon=0.2$ was used. For the calculations
with {\casdk}, we employ the {\small VMEC} tokamak equilibrium described
above, which has similar parameters at $\psi=0.7$. At
  this radial position, the {\small VMEC} equilibrium satisfies $q=2$
  and $\varepsilon=0.2$ within an error of 1.5\% ($\varepsilon=0.197$ and $q=2.03$). The difference between the {\small VMEC} equilibrium and the analytical one is illustrated in figure \ref{fig:bfield}, where we compare the magnetic field strength along a field line for both equilibria. 
In the {\small VMEC} equilibrium, the value of the magnetic field
strength at the magnetic axis is $B_0=1.87$ T. In general, deviations
from circularity are expected in the numerical equilibrium because of effects like the Shafranov shift that are
not taken into account in the analytical equilibrium. These deviations
are smaller for radial positions closer to the center.  The
comparisons for the cases with fully kinetic species (\ref{eq:xck})
and with adiabatic electrons (\ref{eq:xca}) are shown in figures
\ref{fig:XC2} and \ref{fig:XC1}. The agreement is quite good. The fact
that the curves present some differences, especially at short
wavelengths, is not surprising because the equilibria are not
identical.
As already advanced above, we will see that the
{\casdk} calculations agree very well with the results from gyrokinetic 
simulations carried out with \gene~coupled to {\small GIST} \cite{Xanthopoulos2009} and {\small EUTERPE}, that also employ {\small VMEC} equilibria. It should be noted that further gyrokinetic codes with similar capabilities exist. The independently developed code {\small GKV-X} \cite{Nunami2010}, for instance, is also able to handle {\small VMEC} equilibria.

In the next section, we compare {\casdk} calculations of the residual
zonal flow with the results obtained from gyrokinetic simulations.

\section{Comparison of the residual zonal flow values obtained with {\small CAS3D-K} and with gyrokinetic simulations}
\label{sec:results}

In this section, we calculate the residual zonal flow as an initial value problem for a wide range of radial wavelengths in tokamak and stellarator geometries.  We use the numerical techniques explained in Section~\ref{sec:cas3dk} to evaluate the required averages using the code {\casdk}.  These calculations will be compared with the results from two gyrokinetic codes, showing that the extension of {\casdk} is faster.

The residual level for the initial value problem is given by (\ref{eq:rsk}), once an initial condition $f_s(0)$ has been specified. This initial condition has to satisfy (\ref{eq:qn4}). An initial condition fulfilling this equation is
\begin{equation}
 f_s(0) = 
  \frac{Z_s e}{T_s} \frac{ \left\langle
   1-\Gamma_{0s}\right\rangle_\psi }{\Gamma_{0s}} 
  J_{0s} F_{s0}\, \varphi_k(0).
  \label{eq:ic}
\end{equation}
Using (\ref{eq:ic}), we find that the expression for the residual level is 
\begin{equation}
 \frac{\varphi_k(\infty)}{\varphi_k(0)} =
  \frac{
  \sum_s \frac{Z_s^2}{T_s} 
  \left\{ \mathrm{e}^{-\mathrm{i}k_\psi\delta_s}J_{0s} \,
   \overline{\mathrm{e}^{\mathrm{i}k_\psi\delta_s}  
   J_{0s} \left\langle 1-\Gamma_{0s}\right\rangle_\psi
   /\Gamma_{0s} } 
	 \right\}_s^{\overline{\omega_s}=0}
  }{
  \sum_s \frac{Z_s^2}{T_s} 
  \left[\left\{1\right\}_s-\left\{
	 \mathrm{e}^{-\mathrm{i}k_\psi\delta_s} J_{0s} \, 
	 \overline{ \mathrm{e}^{\mathrm{i}k_\psi\delta_s} J_{0s}}
			   \right\}_s^{\overline{\omega_s}=0} 
	\right]
  }.
  \label{eq:rltk}
\end{equation}
Similarly, from (\ref{eq:rsa}) and (\ref{eq:ic}) we obtain the residual level when using the approximation of adiabatic electrons. Namely,
\begin{equation}
 \frac{\varphi_k(\infty)}{\varphi_k(0)} =
  \frac{ \sum_{s\neq e} \frac{Z_s^2}{T_s} 
  \left\{ \mathrm{e}^{-\mathrm{i}k_\psi\delta_s}J_{0s} \,
   \overline{\mathrm{e}^{\mathrm{i}k_\psi\delta_s}  
   J_{0s} \left\langle 1-\Gamma_{0s}\right\rangle_\psi /\Gamma_{0s} }
		\right\}_s^{\overline{\omega_s}=0}
  }{
  \sum_{s\neq e} \frac{Z_s^2}{T_s} 
  \left[\left\{1\right\}_s-\left\{
	 \mathrm{e}^{-\mathrm{i}k_\psi\delta_s} J_{0s} \, 
	 \overline{ \mathrm{e}^{\mathrm{i}k_\psi\delta_s} J_{0s}}
			   \right\}_s^{\overline{\omega_s}=0} 
  \right]
  }.
  \label{eq:rlta}
\end{equation}
As explained above, in our numerical evaluations of (\ref{eq:rltk})
and (\ref{eq:rlta}) with {\casdk}, we assume that in stellarator
geometry the trapped trajectories have $\overline{\omega_s} \neq 0$.

To be precise, the Xiao and Catto formulas
  (\ref{eq:xck}) and (\ref{eq:xca}) can be obtained from an initial
  value problem calculation by choosing an initial condition $f_s(0)$
  different from ours. However, the initial conditions that recover
  the X-C results necessarily have increasingly fast
    oscillations along the orbit for increasing $k_\psi$, and seem of
  limited interest for the analysis of turbulence simulations. For
  this reason, we choose a different initial condition that is in our
  opinion more relevant.

The comparison with gyrokinetic simulations will be carried out by using the \gene~and {\small EUTERPE} codes. \gene~\cite{Jenko2000,Gorler2011,GENE, Xanthopoulos2009} is a Eulerian gyrokinetic $\delta f$ code which can be run in radially global, full flux surface or flux tube simulation domains.  The code can use adiabatic or kinetic electrons and is able to deal with tokamak and stellarator geometries.  In the \gene~simulations, we calculate the zonal flow response for a wide range of radial wavelengths, using both adiabatic and kinetic electrons. {\small EUTERPE}~\cite{Jost2001,Kleiber2012} is a global $\delta f$ gyrokinetic code in 3D geometry with a Lagrangian Particle In Cell (PIC) scheme.  In the simulations with {\small EUTERPE}, a $k_\perp\rho_{ts} < 1$ approximation is employed in the quasineutrality equation that limits the range of wavelengths for which we can carry out the calculations.  With {\small EUTERPE}, we have been able to simulate adiabatic electrons and also kinetic heavy electrons.  
All the gyrokinetic 
simulations shown in this work are linear and collisionless, with a plasma that, unless stated otherwise, consists of only two species: singly charged ions and electrons, $s=\{i,e\}$. We take flat density and temperature profiles with the same values for both species.

In {\small EUTERPE} an initial condition proportional to $\sin(k_\psi\psi)$ is used for the perturbed distribution function. After the first time step, a zonal perturbation to the potential with the same radial dependence $\varphi(0)\propto \sin( k_\psi \psi)$ appears, that is used as the initial zonal flow. For the implementation of the initial condition in {\small EUTERPE}, the Bessel functions $J_{0s}$ and $\Gamma_{0s}$ are approximated to lowest order in $k_\perp\rho_{ts}\ll 1$.  The initial condition in {\small EUTERPE} is then 
\begin{equation}
 F_{s1}(0) = 
  \frac{Z_s e}{T_s} \left\langle k_\perp^2\rho_{ts}^2 
\right\rangle_\psi \varphi_k(0) \sin(k_\psi \psi)\, F_{s0}.
\end{equation} 
Since \gene~works in Fourier space for the radial coordinate, we initialize the perturbed distribution function with only one radial mode which produces a potential with a single mode of unit amplitude.

\begin{figure}
\includegraphics[width=1.\linewidth]{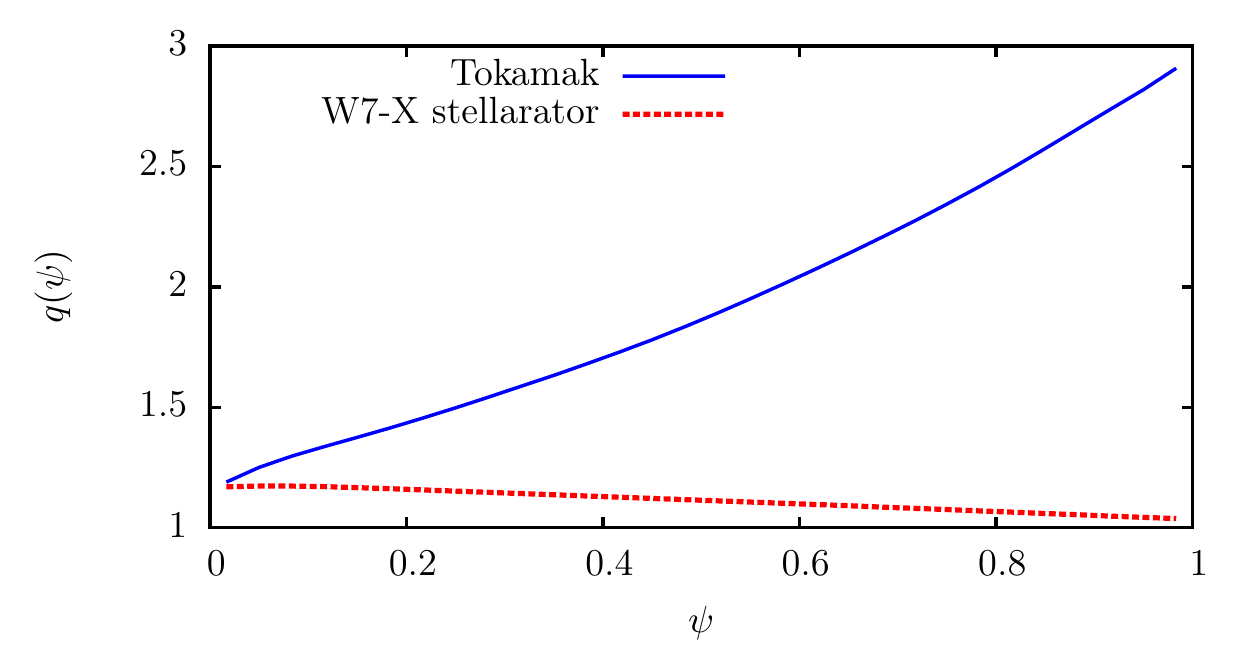}
\caption{Safety factor profiles employed in Section \ref{sec:results} for the tokamak (solid line) and W7-X (dashed line) calculations.}
\label{fig:qprofileSectionGyroSim}
\end{figure}

\subsection{Tokamak}
\label{sec:Tokamakresults}

\begin{figure}
\includegraphics[width=1.\linewidth]{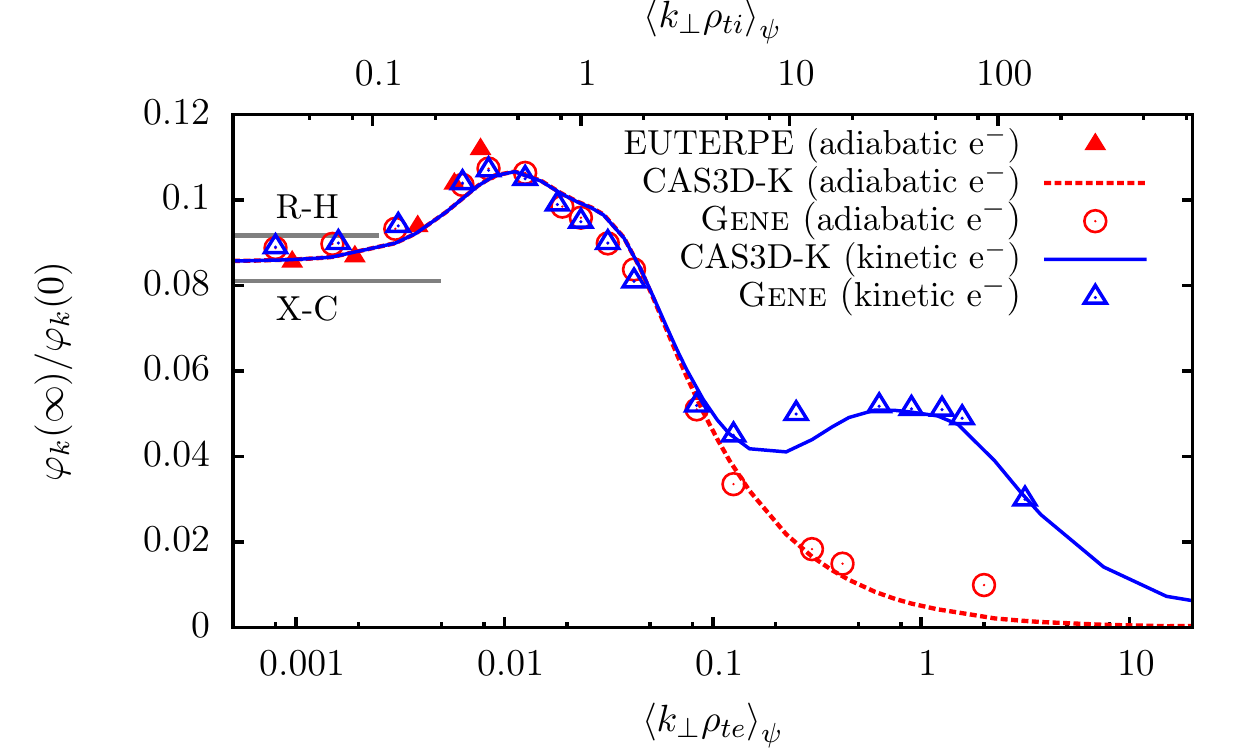}
\caption{Residual zonal flow for the initial value problem in an axisymmetric large aspect ratio tokamak with major radius $R=0.95$~m, minor radius $a=0.25$~m and $q$ profile given in figure \ref{fig:qprofileSectionGyroSim}. The values predicted by R-H and X-C (equations (\ref{eq:rh}) and (\ref{eq:XCkequaltozero}), respectively) are also shown for comparison.}
\label{fig:tk_kae}
\end{figure}

First, we compare gyrokinetic simulations and {\casdk} calculations in tokamak geometry. We use an axisymmetric device with major radius $R=0.95$~m, minor radius $a=0.25$~m, and $q$ profile given in figure \ref{fig:qprofileSectionGyroSim}, whose equilibrium is determined by {\small VMEC}. We use flat temperature profiles with $T_i=T_e$. The residual levels obtained with {\small EUTERPE}, {\casdk} and the flux tube version of \gene~are shown in figure \ref{fig:tk_kae} for a radial position $\psi=0.25$. We show the calculations with fully kinetic species and also using the approximation of adiabatic electrons. The results of the gyrokinetic codes have been obtained fitting the temporal evolution of the potential to an exponential decay model 
\begin{equation}
 \varphi(t)/\varphi(0) = 
  R + A \exp({-\xi t}).
  \label{eq:expfit}
\end{equation}
The results with {\casdk} correspond to the evaluation of the equations (\ref{eq:rltk}) and (\ref{eq:rlta}). From figure \ref{fig:tk_kae}, we can see that the agreement among the results of {\casdk}, {\small EUTERPE} and \gene~is excellent. When evaluating the residual level with gyrokinetic codes, a certain variability in the results must be assumed. This variability comes from the fitting method (smaller than 1\%), the discretization in phase space and the control of the numerical noise, among other factors. All the results in this work obtained with {\gene} show variations smaller than 10\%. In any case, figure \ref{fig:tk_kae} shows that the residual level obtained by the three independent methods coincides within a margin smaller than this quantity, which gives us confidence to consider that the overall error is quite small. 

\begin{figure}
\includegraphics[width=1.\linewidth]{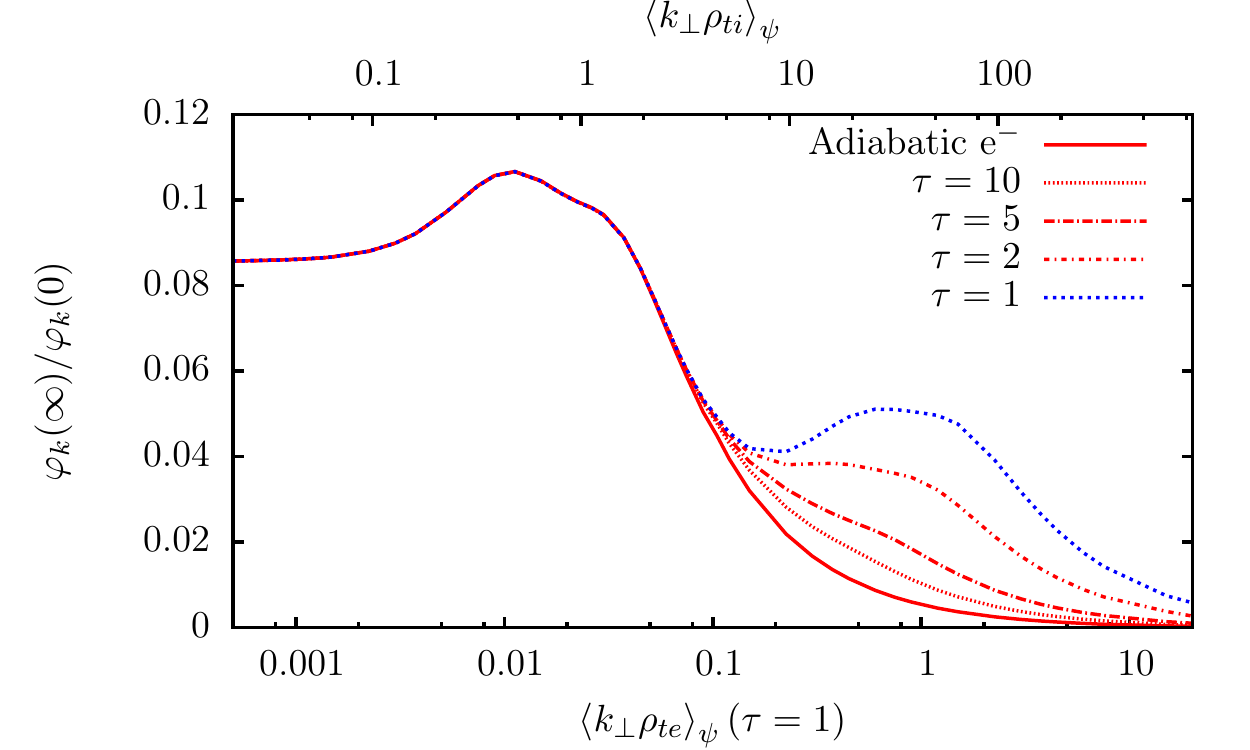}
\caption{The same evaluation with {\footnotesize CAS3D-K} of equation (\ref{eq:rlta}) as in figure \ref{fig:tk_kae}, with the adiabatic electron approximation, and the evaluation of equation (\ref{eq:rltk}) with fully kinetic species for different values of $\tau$.} 
\label{fig:larttau}
\end{figure}

As can be seen in figure \ref{fig:tk_kae}, the residual value has local maxima centered at the scales of the electron and ion Larmor radii.  In the long-wavelength limit, $k_\perp\rho_{ti} \ll 1$, the residual level in a tokamak does not depend on $k_\perp$. Its value is well predicted by (\ref{eq:rh}), and even more accurately by (\ref{eq:XCkequaltozero}), for a large aspect ratio tokamak with circular cross section. For the {\small VMEC} equilibrium used here, these predictions (also indicated in figure \ref{fig:tk_kae}) are not so accurate for the reasons discussed in the paragraph below equation (\ref{eq:XCkequaltozero}). At very short wavelengths, $k_\perp\rho_{te}>2$, the residual value approaches zero as $\left\langle k_\perp \rho_{ts} \right\rangle_\psi^{-1}$ when using fully kinetic species and also for the adiabatic electron approximation. In figure \ref{fig:tk_kae}, it is shown that the adiabatic electron approximation in tokamaks is good for $k_\perp\rho_{te} \lesssim 0.1$. In figure \ref{fig:larttau}, we 
reproduce the results with {\casdk} in figure \ref{fig:tk_kae} together with the evaluation of (\ref{eq:rltk}) for different values of $\tau:=T_e/T_i$.  On the electron scale, the results with the adiabatic electron approximation and with kinetic electrons only coincide in the limit $\tau\gg1$.  Due to the reasons pointed out above, the simulations with {\small EUTERPE} have been carried out only for $k_\perp\rho_{ti}< 1$.  Finally, it is obvious that the results of the forced system of figures \ref{fig:XC2} and \ref{fig:XC1} and the initial value problem of figure \ref{fig:tk_kae} behave in a completely different way for $k_\perp\rho_{ti}\gtrsim 1$.

\subsection{Stellarator}
\label{sec:Stellaratorresults}

\begin{figure}
\includegraphics[width=1.\linewidth]{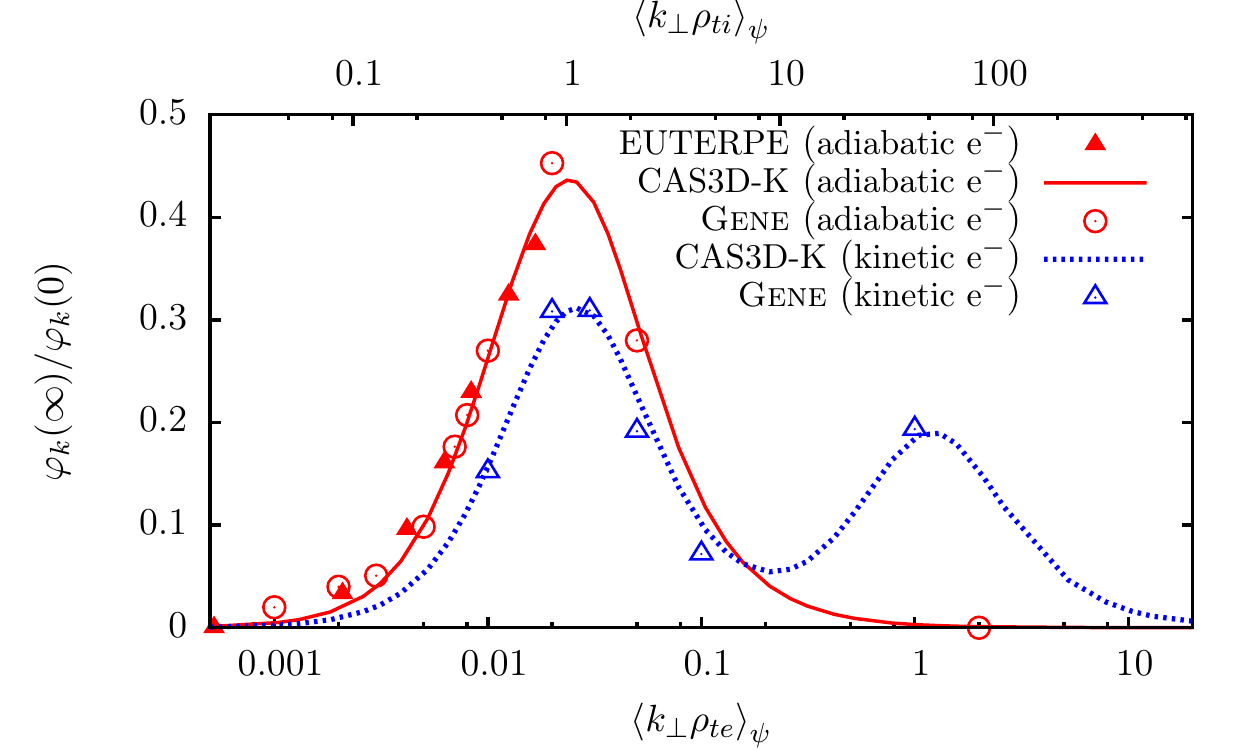}
\caption{Residual zonal flow level for the initial value problem in the standard configuration of the W7-X stellarator at $\psi=0.25$. The $q$ profile is shown in figure \ref{fig:qprofileSectionGyroSim}.}
\label{fig:W7Xak}
\end{figure}

Now, we turn to stellarator geometry. We use an equilibrium for the standard configuration of the stellarator W7-X obtained with {\small VMEC}. The $q$ profile is given in figure \ref{fig:qprofileSectionGyroSim} and we take flat density and temperature profiles with $T_i=T_e$.  In figure \ref{fig:W7Xak}, calculations of the residual level with {\casdk}, {\small EUTERPE} and the full flux surface version of \gene~are shown for $\psi=0.25$.  Two curves correspond to {\casdk} computations, one using adiabatic electrons (see equation (\ref{eq:rlta})) and the other one using kinetic electrons (see equation (\ref{eq:rltk})). In figure \ref{fig:W7Xak}, the results of the gyrokinetic simulations were obtained employing both the approximation of adiabatic electrons and fully kinetic species with \gene, whereas only calculations with adiabatic electrons are shown for {\small EUTERPE}. These results have been fitted to an exponential decay model (\ref{eq:expfit}) to get the residual value. Similar results can be obtained 
with an algebraic decay model as suggested in reference \cite{Helander2011}. The results of {\casdk} show remarkable agreement with both gyrokinetic codes.

\begin{figure}
\includegraphics[width=1.\linewidth]{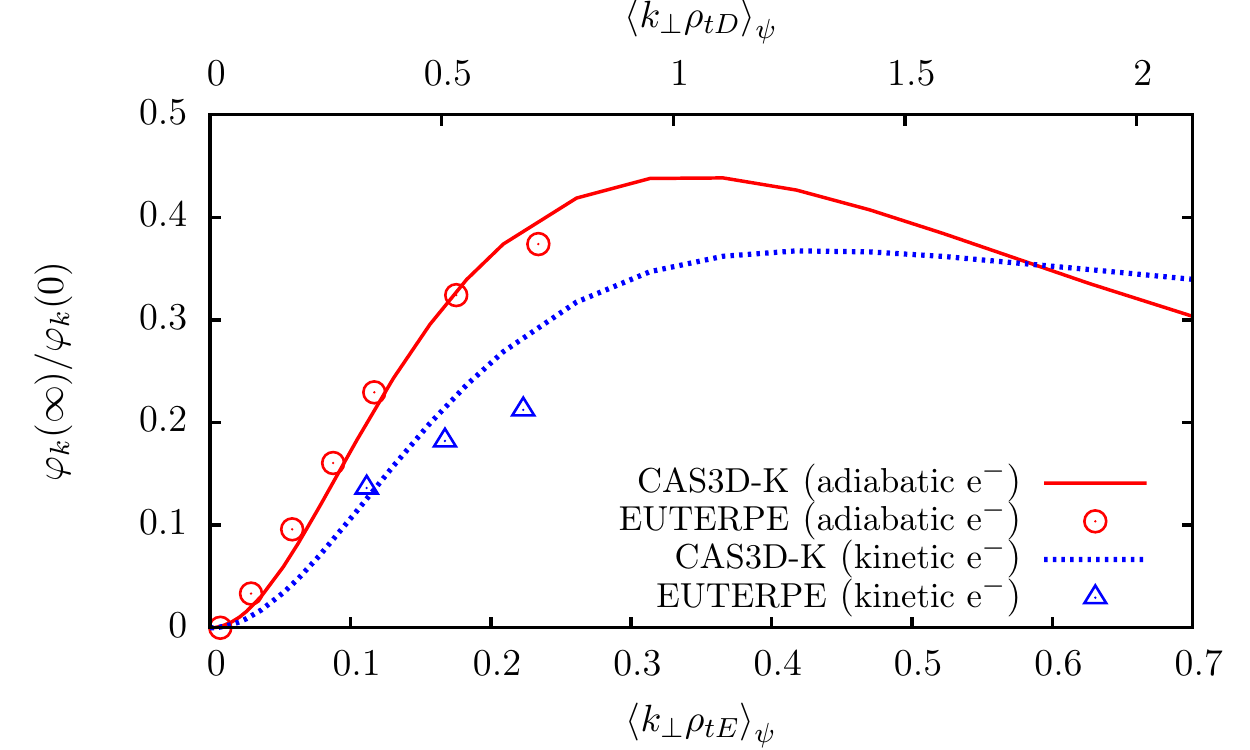}
\caption{Residual zonal flow level for the initial value problem in the standard configuration of the W7-X stellarator at $\psi=0.25$, with deuterium ions ($D$) and kinetic heavy electrons ($E$) ($m_E = 400 m_e$) and also using the approximation of adiabatic electrons.}  
\label{fig:w7x-e_fatD-euterpe-lin-8}
\end{figure}

As explained and quantified at the end of this section, the gyrokinetic simulations with kinetic species are much more demanding in terms of computational resources than those with adiabatic electrons. Global simulations with {\small EUTERPE} using fully kinetic electrons in stellarator geometry would require an extremely large computing time. This time can be reduced by increasing the mass of the species involved.  We have calculated for deuterium ions and kinetic heavy electrons $s=\{D,E\}$, with $m_E = 400\,m_e$ and $T_D=T_E$.  The results are shown in figure \ref{fig:w7x-e_fatD-euterpe-lin-8} where we compare the residual level calculated with {\small EUTERPE} and {\casdk} at $\psi=0.25$.  The results with adiabatic electrons shown in this figure are exactly the same as those with adiabatic electrons in figure \ref{fig:W7Xak}, obtained for hydrogen ions.  Note that, with adiabatic electrons, as the residual level only depends on $\left\langle k_\perp\rho_{ti} \right\rangle_\psi$, the 
curves for hydrogen or deuterium ions are exactly the same.

\begin{figure}
\includegraphics[width=1.\linewidth]{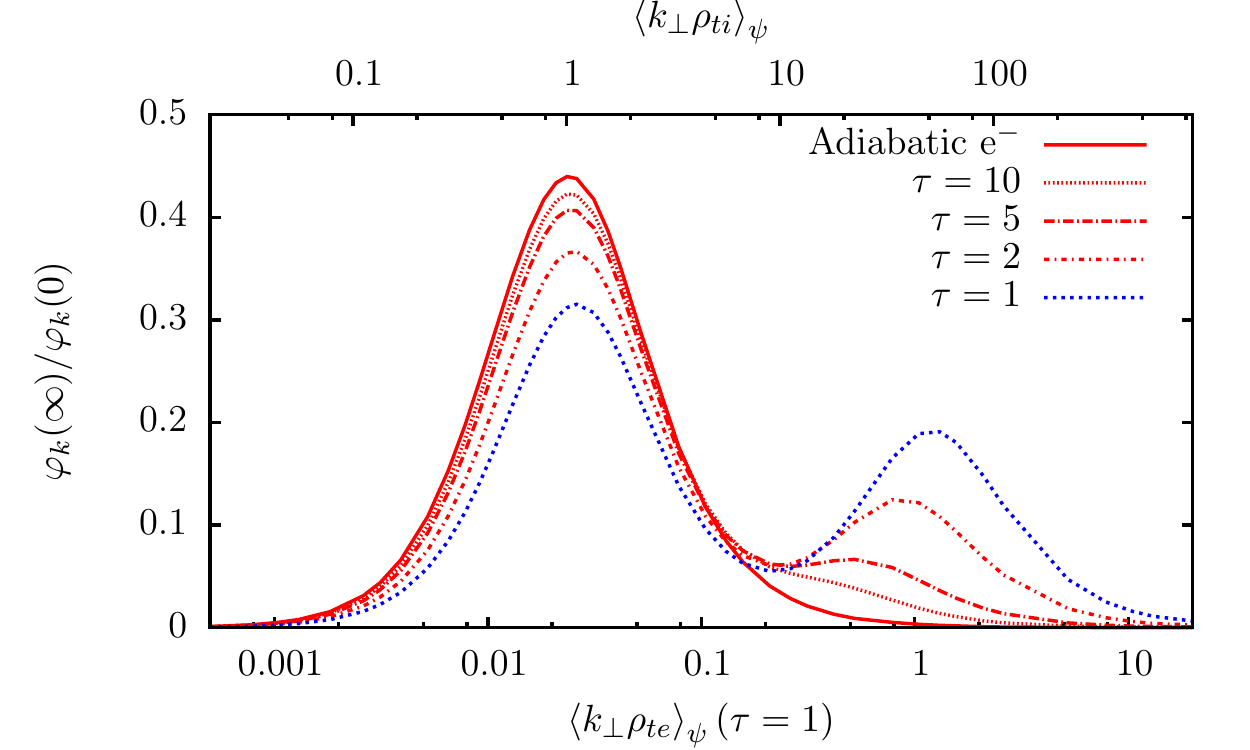}
\caption{The same evaluation with {\footnotesize CAS3D-K} of equation (\ref{eq:rlta}) as in figure \ref{fig:W7Xak}, with the adiabatic electron approximation, and the evaluation of equation (\ref{eq:rltk}) with fully kinetic species for different values of $\tau$.} 
\label{fig:w7xtau}
\end{figure}

In figure \ref{fig:W7Xak}, like in tokamaks, we find local maxima of the residual level centered around the scales of the electron and ion Larmor radii. However, at long wavelengths, $k_\perp\rho_{ti}\ll 1$, the residual level as a function of $k_\perp \rho_{ti}$ behaves very differently in tokamaks and in stellarators (see, for example, figures \ref{fig:tk_kae} and \ref{fig:W7Xak}). This can be easily understood by expanding (\ref{eq:rltk}) and (\ref{eq:rlta}) in $k_\perp\rho_{ti}\ll 1$. The numerator of these expressions scales quadratically with $k_\perp \rho_{ti}$ in tokamaks and stellarators. The difference comes from the denominator. In a stellarator, the denominator is non-zero when $k_\perp\rho_{ti} = 0$. However, in a tokamak the denominator scales quadratically with $k_\perp\rho_{ti}$. The denominator has been often related to the shielding effects of collisionless classical and neoclassical polarization currents \cite{Rosenbluth1998, Watanabe2008, Xiao2006, Xiao2007}. 

\begin{figure}
\includegraphics[width=1.\linewidth]{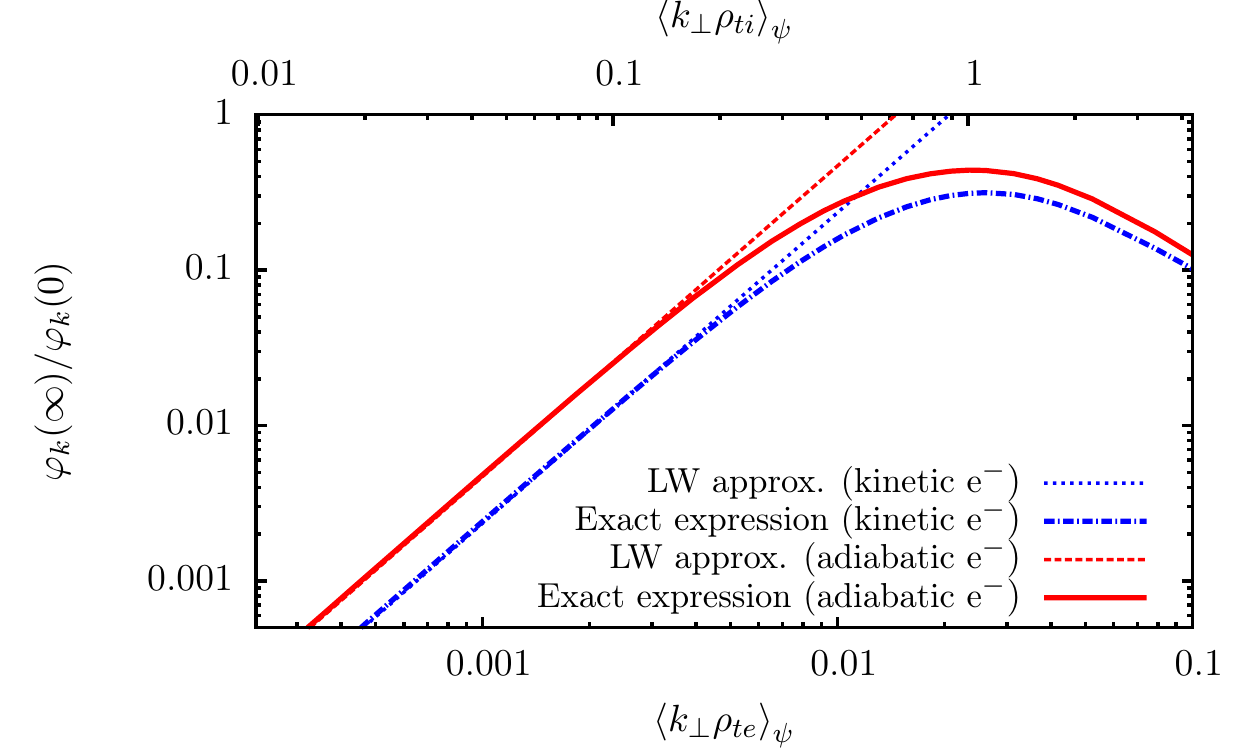}
\caption{Range of validity of the long-wavelength (LW) approximations, when using fully kinetic species (\ref{eq:stelllwke}) and with the approximation of adiabatic electrons (\ref{eq:stelllwae}), compared to the exact expressions (\ref{eq:rltk}) and (\ref{eq:rlta}), in the standard configuration of the W7-X stellarator at $\psi=0.25$ and with $T_i=T_e$.} 
\label{fig:lwapprox}
\end{figure}

It is worth giving explicitly the $k_\perp\rho_{ti}\ll 1$ expansions of (\ref{eq:rltk}) and (\ref{eq:rlta}) in a stellarator and discussing a stellarator specific point in detail. The lowest order term of (\ref{eq:rltk}) gives 
\begin{equation}
  \frac{\varphi_k(\infty)}{\varphi_k(0)}=
\frac{(1-\epsilon_t)\left\langle k_\perp \rho_{ti} 
    \right\rangle_\psi^{2}}{\epsilon_t
    \left(1+T_i/(Z_i^2T_e)\right)} +O(\left\langle k_\perp \rho_{ti} 
    \right\rangle_\psi^{4}),
\label{eq:stelllwke}
\end{equation}
where $\epsilon_t = n_s^{-1}\{1\}_s^{{\rm trapped}}$ is the fraction of trapped particles. Here, the superindex ``trapped'' means that the phase space integration is performed only over the trapped region. However, if we use the approximation of adiabatic electrons, (\ref{eq:rlta}), we find 
\begin{equation}
  \frac{\varphi_k(\infty)}{\varphi_k(0)}=
  \frac{(1-\epsilon_t)\left\langle k_\perp \rho_{ti} 
    \right\rangle_\psi^{2}}{\epsilon_t}+O(\left\langle k_\perp \rho_{ti} 
  \right\rangle_\psi^{4}).
\label{eq:stelllwae}
\end{equation}
Hence, in stellarators, the adiabatic electron approximation gives and incorrect residual zonal flow, even at long wavelengths.  This has been pointed out in references~\cite{Sugama2005, Sugama2006} and is confirmed by the calculations shown in figures \ref{fig:W7Xak} and \ref{fig:w7x-e_fatD-euterpe-lin-8}.  The curves in figure \ref{fig:w7xtau} for different values of $\tau$ quantify the error of the adiabatic electron approximation for any wavelength.  As can be seen in this figure, the residual level obtained with this approximation, only coincides with that obtained with fully kinetic species in the limit $\tau\gg1$.  In figure \ref{fig:lwapprox}, we plot the curves in figure \ref{fig:W7Xak} corresponding to {\casdk} together with the evaluation of their expansions to lowest order in $k_\perp\rho_{ti}\ll 1$, (\ref{eq:stelllwke}) and (\ref{eq:stelllwae}).  It is clear from figure~\ref{fig:lwapprox} that (\ref{eq:stelllwke}) and (\ref{eq:stelllwae}) are good approximations of (\ref{eq:rltk}) and (\ref{eq:rlta}) respectively for $k_\perp\rho_{ti} \lesssim 0.2$.

We point out that at scales comparable to the ion Larmor radius,
$k_\perp\rho_{ti}\sim 1$, the residual level appears to be larger in
stellarators than in tokamaks (see, for example, figures
\ref{fig:larttau} and \ref{fig:w7xtau}). In order to discard trivial
explanations, we have studied with {\casdk} the residual level in a
tokamak configuration with the same aspect ratio and $q$ profile as
those of the standard configuration of W7-X and the results are much
closer to the tokamak case than to the stellarator results. This is
not surprising because, as shown in reference \cite{Xiao2007}, not
only the aspect ratio but also shaping effects like elongation,
triangularity and Shafranov shift, among others, affect the residual
level. We leave for a future work a detailed study of the magnetic
configuration influence on the residual level.

\subsection{Simulation conditions and computational time}
\label{sec:ComputationalTime}

The converged results shown in this work require disparate computing
resources, depending on the code and the physical problem.  
The relevant numerical parameters and the computational
  resources required by each code are described
  below.

In {\casdk}, we used 256 points for the integration over the velocity coordinate $v$, with $0\leq v\leq 4\pi v_{ts}$. For the integration over the $\lambda$ coordinate, we used 72 integration points in the tokamak and 24 in W7-X. 
Along the field line, we used 32 points at long wavelengths and up to 4096 for short wavelengths. This resolution allows the correct integration of the highly oscillatory functions. Thanks to axisymmetry, in a tokamak all field lines on a flux surface are equivalent. In W7-X, we used 1024 field lines to cover the flux surface for passing particles. For the evaluation of $\delta_s$ in the stellarator case, all the modes with $|m|\leq 8$ and $|n|\leq 8$ were retained. The calculations were carried out in the EULER cluster at CIEMAT, equipped with Xeon 5450 quadcore processors at 3 GHz and 4XDDR Infiniband network.

In \gene, a 1D spatial grid
  along the field line ($z$ coordinate) is used in the tokamak cases
  while in stellarator simulations a 2D spatial grid in coordinates $(y,z)$ is used to describe a full flux surface ($y$ is the
  coordinate along the binormal direction). In
    velocity space, a 2D grid in parallel velocity and magnetic
  moment coordinates $(v_{\parallel}, \mu)$ is used in both the
  tokamak and the stellarator cases.  The resolution of the spatial
  and velocity grids used are given in table \ref{tab:GENE} together
  with the time step and the total simulation time for each case. Times are given in $1/\Omega_G$ units, with $\Omega_G=a/v_{te}$, where we recall that $a$ is the minor radius and $v_{te}$ is the thermal velocity of electrons. 
The GENE simulations were run in HYDRA \cite{HYDRA}, equipped with Intel Ivybridge at 2.8 GHz and SandyBridge-EP  at 2.6 GHz processors interconnected by Infiniband FDR14.

\begin{table}
\centering
\begin{tabular}{ r ccccc c}
\hline \hline
 &\multicolumn{2}{c}{long wavelength ($k_\perp\rho_{ti}< 1$)} & &  \multicolumn{2}{c}{short wavelength ($k_\perp\rho_{ti}> 1$)} &\\
 \cline{2-3} \cline{5-6} & adiabatic e$^{-}$ & kinetic e$^{-}$ & &  adiabatic e$^{-}$ & kinetic e$^{-}$ &\\ 
\hline 
$n_z $ & $64$ & $64$ & & $128$ & $256$ &\\
$n_{v_{\parallel}} $ & $128$ & $1024$ & & $1024-2048$ & $2048 - 4096$\\
$n_{\mu}$ & $40$ & $40$ &   & $40$  & $ 40$
&\hspace{-.5cm} tokamak \\
$\Delta t_G$ & $1-6$ & $0.03-0.06$ & & $0.2-1$ & $0.02$& \\
$T_G$ & $4000-15000$ & $3000-10000$ & & $4000-10000$ & $150$ & \\
\hline
$n_y$ & $64$ &  $ 64$& &    $ 64$ &  $ 64$  & \\
$n_z $ & $256$ & $128$ & & $128$ & $128$ & \\
$n_{v_{\parallel}}$ & $128$ & $256-512$ & & $128$ & $128 -512$ & \hspace{-.5cm} stellarator \\
$n_{\mu}$ & $20$ & $20$ & & $20$ & $10-20$ & \hspace{-.5cm}  \\
$\Delta t_G$ & $4-8$ & $0.06$ & & $0.2-5$ & $0.06$ &\\
$T_G$ & $100000-200000$ & $40000$ & & $300-5000$ & $10000-550000$ &\\
 \hline \hline
\end{tabular}
\caption{Numerical parameters used in GENE simulations. The time step ($\Delta t_G$) and the total simulation time ($T_G$) are given in $\Omega_G^{-1}$ units, with $\Omega_G=a/v_{te}$.}
\label{tab:GENE}
\end{table}

In {\small EUTERPE}, the electric potential is represented in a 3D spatial grid in PEST coordinates $(s_E,\theta_E,\phi_E)$ whose radial resolution must be large enough to correctly represent the potential perturbation. The number of markers was set according to the grid resolution to maintain the ratio of markers per grid cell approximately constant. A low-pass squared filter in Fourier space ($k_{\theta_E},k_{\phi_E}$) is used to reduce the noise.
 In table \ref{tab:EUTERPE} the resolution of the spatial grid $(n_{s_E},n_{\phi_E}, n_{\phi_E})$, the number of markers, the filter cutoff,  the time step and the total simulation time used for each case are given. Times are given in $1/\Omega_E$ units, where $\Omega_E=eB^*/m_i$, $e$ is the elementary charge, $m_i$ is the ion mass and $B^*$ is the average of the magnetic field along the magnetic axis.
The {\small EUTERPE} simulations were carried out in EULER and MareNostrum III \cite{MN3}, equipped with Intel SandyBridge-EP processors at 2.6 GHz and Infiniband FDR10 interconnection.

\begin{table}
\centering
\begin{tabular}{ r ccccc}
\hline \hline
 {\centering long wavelength}&\multicolumn{2}{c}{tokamak} & & \multicolumn{2}{c}{stellarator}\\
\cline{2-3} \cline{5-6}($k_\perp\rho_{ti}< 1   $) & adiabatic e$^{-}$  &  kinetic e$^{-}$&  &adiabatic e$^{-}$ & kinetic e$^{-} {^\dag}$  \\ 
\hline \hline 
$n_{s_E}$ & $32-192$ & ---  & & $64-192 $ & $32-96$ \\
$n_{\theta_E} \times n_{\phi_E}$ &$16 \times 16$ & --- & &  $32 \times 32$ & $16 \times 16$   \\
\# of markers & 40 M $-$ 240 M & --- &  & 40 M $-$ 240 M & $40$ M $-$ 120 M  \\
filter cutoff& $5 $ & --- &  & $5, 10 ^{\dag\dag}$ & $5 $   \\
$\Delta t_E $ & $10 - 5$& --- &  & $50 - 20 $ & $0.5$  \\
$T_E$ & $60000$&  --- &  & $400000$ & $45000$    \\
 \hline \hline
\end{tabular}
\caption{Numerical parameters used in EUTERPE simulations. The time step ($\Delta t_E$) and the total simulation time ($T_E$) are given in $\Omega_E^{-1}$ units, with $\Omega_E=eB^*/m$. $^\dag$This range corresponds to calculations with deuterium ions and heavy electrons. $^{\dag\dag}$Only for the shortest-wavelength case.}
\label{tab:EUTERPE}
\end{table}

In table~\ref{tab:cpuhts} we illustrate
  the computational cost, in total CPU core hours (that
  is, the time summed up over all the cores employed in the
  simulation), for the different codes and cases studied.  Of course,
  the values shown in table~\ref{tab:cpuhts} are
  simply indicative, as they depend on the numerical
  details of the simulations and the type of CPU employed in each
  calculation. In addition, a systematic analysis
  of the optimal resolution to carry out the computations with each
  code has not been performed. The main conclusion that we can extract
  from table~\ref{tab:cpuhts} is that determining the residual zonal flow with
  {\casdk} is faster than with {\gene} and {\small EUTERPE}. This is
  specially true for stellarators. The reason is that in
stellarators only passing particles contribute to the residual value,
while in tokamaks also trapped particles count, and trapped
trajectories typically demand a more careful numerical treatment than
passing ones. Whereas the {\casdk} calculation simply drops the
contribution from the trapped region, the gyrokinetic runs simulate
all trajectories. However, we can see from table~\ref{tab:cpuhts} that
the computational cost when using the gyrokinetic
codes is higher in stellarator geometry because it requires increased
resolution in phase space to obtain converged results.

\begin{table}
\centering
\begin{tabular}{ r ccccc c}
\hline \hline
 &\multicolumn{2}{c}{long wavelength ($k_\perp\rho_{ti}< 1$)} & & \multicolumn{2}{c}{short wavelength ($k_\perp\rho_{ti}> 1$)} &\\
 \cline{2-3} \cline{5-6} & adiabatic e$^{-}$ & kinetic e$^{-}$ & & adiabatic e$^{-}$ & kinetic e$^{-}$ &\\ 
\hline 
{\casdk} & $ 1.5$ & $ 3$ & & $ 15$  &  $ 30$ &\\
\gene & $10^{\dag\dag}$ & $ 1300$ & &  $ 150$ & $ 200$ &\hspace{-.5cm} tokamak \\ 
{\small EUTERPE} & $ 7000$ & --- & & --- & --- &\\
\hline
{\casdk} & $ 0.5$ & $ 1$ & & $ 4$ & $ 8$ &\\
\gene & $ 2000$ & $ 40000$ &  & $ 200$  & $ 250000$  &\hspace{-.5cm} stellarator \\
{\small EUTERPE} & $20000$ & $80000^\dag$ & & --- & ---&  \\
 \hline \hline
\end{tabular}
\caption{Estimated CPU time (total core hours) to obtain the residual zonal flow value with the different codes. We give estimations for tokamaks and stellarators; for adiabatic and kinetic electrons; and for long and short wavelengths. $^\dag$This range corresponds to calculations with deuterium ions and heavy electrons. $^{\dag\dag}$For very small wavenumbers, \gene~computes the residual zonal flow in a tokamak in approximately 0.5 CPU hours.}
\label{tab:cpuhts}
\end{table}

We observe that {\small EUTERPE}, a 3D global code, requires much more
CPU time than \gene, particularly in the tokamak case, in which the
flux tube version of \gene~is used. The reason is that {\small
  EUTERPE} simulates the whole plasma while \gene~is here operated in
a radially local limit. The computational cost with
  {\small EUTERPE} increases with $k_\perp\rho_{ti}$, because more
flux surfaces have to be considered as $k_\perp$ increases to properly
resolve the radial structure of the potential in all the plasma
volume. In {\small EUTERPE} the different values of $k_\perp\rho_{ti}$
at a given radial position are obtained by keeping the value of
$\rho_{ti}$ (determined by the ion mass, the temperature and the
magnetic field) and varying the value of $k_\perp$.  In {\casdk}, the
resolution at short wavelengths must be increased to correctly
calculate the highly oscillatory functions related to the finite orbit
width and the finite Larmor radius effects.

The analytical expression obtained by Rosenbluth and Hinton in
reference \cite{Rosenbluth1998}, given by equation (\ref{eq:rh}), has
been largely used as a linear benchmark for gyrokinetic codes in
tokamak geometry and in the long-wavelength limit. The results
presented in this work show that {\casdk} can be used to perform those
benchmarks not only in tokamak geometry but also in stellarator
geometry and for arbitrary wavelengths. Examples of such benchmarks
are given in figure~\ref{fig:tk_kae} for the global code {\small
  EUTERPE} and the flux tube version of \gene~in tokamak geometry, and
in figure~\ref{fig:W7Xak} for {\small EUTERPE} and the full flux
surface version of \gene~in stellarator geometry.

Finally, it is a matter of fact that the tokamak calculation including
a source term in the quasineutrality equation~\cite{Xiao2006,Xiao2007}
has become quite popular in the literature. Just for completeness, we
give an analogous calculation for the stellarator in \ref{sec:forced}
using gyrokinetic simulations and {\casdk}. We also show the results for
the tokamak using gyrokinetic codes (these were not included in
Section \ref{sec:cas3dk}).

\section{Conclusions}
\label{sec:conclusions}

In this work we have treated analytically the linear collisionless zonal flow evolution as an initial value problem, and derived expressions (see (\ref{eq:rsk}) and (\ref{eq:rsa})) for the residual value that are valid for arbitrary wavelengths and for tokamak and stellarator geometries. The expressions (\ref{eq:rsk}) and (\ref{eq:rsa}) involve certain averages in phase space, and also the solution of magnetic differential equations, that cannot be evaluated analytically except for very special situations.  We have extended the code {\casdk} to evaluate such expressions in general. We have tested the extension of the code by comparing its results with analytical formulae available in the literature for simplified tokamak geometry~\cite{Xiao2006, Xiao2007}. These tests are given in figures~\ref{fig:RHXC}, \ref{fig:XC2} and \ref{fig:XC1}. 

Then, we have computed the residual zonal flow level in tokamak and stellarator geometries for a wide range of radial wavelengths, using both the approximation of adiabatic electrons and fully kinetic electrons. We have compared the results of {\casdk} with those obtained from two gyrokinetic codes: the global code {\small EUTERPE} and the radially local versions of \gene~(full flux surface and flux tube). The comparisons are shown in figures~\ref{fig:tk_kae}, \ref{fig:W7Xak} and \ref{fig:w7x-e_fatD-euterpe-lin-8}.
  
A stellarator specific effect has been discussed in detail. Namely, the fact that the adiabatic electron approximation gives  incorrect zonal flow residuals even for $k_\perp\rho_{ti}\ll 1$, unlike in tokamaks. This effect has also been confirmed by means of gyrokinetic simulations.  This is shown in figures \ref{fig:W7Xak}, \ref{fig:w7x-e_fatD-euterpe-lin-8} and \ref{fig:w7xtau}.

Finally, we stress the efficiency of our method to determine the
residual zonal flow. Gyrokinetic simulations with
  \gene~and {\small EUTERPE} to obtain the residual level are
computationally expensive, especially with fully kinetic species, in
the short-wavelength region, and in stellarator geometry. On the
contrary, the calculations with {\casdk} are less
  demanding (see table~\ref{tab:cpuhts}), particularly in stellarator
  geometry. These results show that {\casdk} is a useful tool to
calculate fast and accurately the residual level in any toroidal
geometry and for arbitrary wavelengths. This code is even more useful
in stellarator geometry as kinetic electrons must be considered to
correctly calculate the residual level. It can also provide a good
benchmark for gyrokinetic codes.

\section*{Acknowledgments}
\label{sec:acknowledgements}

The authors thank Peter Catto, Bill Dorland, Per Helander, Alexey Mishchenko and Yong Xiao for helpful discussions, and Antonio L\'opez-Fraguas for his help with the usage of {\small VMEC}. The authors thankfully acknowledge the computer resources, technical expertise and assistance provided by the Barcelona Supercomputing Center (BSC), the Rechenzentrum Garching (RZG) and the Computing Center of
CIEMAT. 

This research has been funded in part by grant ENE2012-30832, Ministerio de Econom\'ia y Competitividad (Spain) and by an FPI-CIEMAT PhD fellowship. This work has been carried out within the framework of the EUROfusion Consortium and has received funding from the Euratom research and training programme 2014-2018 under grant agreement No 633053. The views and opinions expressed herein do not necessarily reflect those of the European Commission.

\appendix

\section{Magnetic differential equation}
\label{sec:mde}

In this section, we give the solution $\delta_s$ to the magnetic differential equation (\ref{eq:mde}). We use Boozer coordinates. That is, we assume that $\{\psi,\theta,\zeta\}$ are such that the contravariant form of $\mathbf{B}$ is given by (\ref{eq:Bcontravariantform}), and its covariant form is given by
\begin{equation}
 {\bf B}
  = I_t \nabla \theta 
  - I_p \nabla\zeta
  + \widetilde{\beta} \left(\psi,\theta,\zeta\right)\nabla\psi.
\end{equation}
The square root of the metric determinant is $\sqrt{g} = (I_t\Psi_p' - I_p \Psi_t')/B^2$.  Here, $I_t=I_t(\psi)$ and $I_p=I_p(\psi)$ are the toroidal and poloidal currents, respectively. In Boozer coordinates, the radial magnetic drift reads
\begin{equation}\label{eq:radialMagDriftBoozer}
\omega_s = -\frac{1}{\tau_b\Psi'_p}
\left[ I_p\partial_\theta + I_t\partial_\zeta \right]\rho_{\parallel s}.
\end{equation}
The parallel gyroradius is defined as $\rho_{\parallel s}=v_{\parallel}/\Omega_s$, whereas $\tau_b = B\sqrt{g}/(v_{\parallel}\Psi'_p)$. 

\subsection{Magnetic differential equation for passing particles}

For passing particles, $\overline{\omega_s}=0$. The magnetic differential equation to be solved is then given by
\begin{equation}
 v_\parallel \hat\mathbf{b}\cdot \nabla \delta_s
  = \omega_s,
  \label{eq:mdep}
\end{equation}
which in Boozer coordinates becomes
\begin{equation}
 \left( \Psi_p' \partial_\theta + \Psi_t' \partial_\zeta \right) \delta_s
 = -\left( I_p \partial_\theta 
     + I_t \partial_\zeta \right)  \rho_{\parallel s}.
\end{equation}
This equation is easily solved in Fourier space, giving
\begin{equation}
 \delta_s 
  = C
  - \sum_{m,n\neq0}
  \left[ \frac{m I_p + n I_t}{m \Psi'_p + n \Psi'_t} \right]
  {(\rho_{\parallel s})}_{mn} \mathrm{e}^{ 2\pi \mathrm{i}(m\theta+n\zeta)},
  \label{eq:mdep1}
\end{equation}
where the coefficients $(\rho_{\parallel s})_{mn}$ are defined by
\begin{equation}
\rho_{\parallel s} = \sum_{m,n} 
 (\rho_{\parallel s})_{mn}
\mathrm{e}^{2\pi \mathrm{i}(m\theta+n\zeta)} 
\end{equation}
and $C$ is a constant. By choosing the integration constant as $C = -I_p(\rho_{\parallel s})_{00}/\Psi'_p$, we have
\begin{equation}
\fl
\delta_s 
  = -\frac{I_p}{\Psi'_p} \left[\rho_{\parallel s}
  + \left(\frac{I_t}{q I_p}-1\right) \sum_{m,n\neq0}
  \left( \frac{qn}{m + qn} \right)
  {(\rho_{\parallel s})}_{mn} \mathrm{e}^{ 2\pi
  \mathrm{i}(m\theta+n\zeta)}\right]. 
  \label{eq:deltap}
\end{equation}
This solution, already given in \cite{Mishchenko2008}, is valid for any toroidal geometry. For axisymmetric tokamaks, it reduces to $\delta_s = -I_p \, \rho_{\parallel s}/\Psi'_p$.

\subsection{Magnetic differential equation for trapped particles}

For trapped particles, the magnetic differential equation is given by 
\begin{equation}
 v_\parallel \hat\mathbf{b}\cdot \nabla \delta_s
  = \omega_s - \overline{\omega_s}.
  \label{eq:mdet}
\end{equation}
In coordinates $\{\psi,\theta,\alpha\}$, with $\alpha = \zeta-q\theta$ (and $\{\psi,\theta,\zeta\}$ being Boozer coordinates), one has
\begin{equation}
 \omega_s 
  = -  \frac{1}{\tau_b\Psi'_p}
            \left[ I_p \partial_\theta   
	     + (I_t-qI_p) \partial_\alpha \right] 
	    \rho_{\parallel s},
\end{equation}
and equation (\ref{eq:mdet}) then reads
\begin{equation}
 \tau_b^{-1} \partial_\theta
  \left(\frac{I_p}{\Psi'_p}\rho_{\parallel s} + \delta_s\right) =
\widetilde{\omega}_{s\alpha},
  \label{eq:mdet1}
\end{equation}
where
\begin{equation}
\widetilde{\omega}_{s\alpha} :=
 \frac{qI_p-I_t}{\Psi_p'} \left[ \tau_b^{-1} \partial_\alpha\rho_{\parallel s}
 - \overline{\tau_b^{-1} \partial_\alpha\rho_{\parallel s}} \right].
\end{equation}
The coordinate along the magnetic field line, $\theta$, is not monotonic over the periodic orbit delimited by the bounce points $\theta_{b1}$ and $\theta_{b2}$. We define a monotonic coordinate $\tau$ by
\begin{equation}
 \tau :=
  \left\{
   \begin{array}{lcr}
    \int_{\theta_{b1}}^{\theta} |\tau_b|\, \mathrm{d}\theta'
     & \mathrm{when}\quad \sigma > 0 \vspace{0.2cm} \\
    \widehat{\tau}_b/2 - \int_{\theta_{b2}}^{\theta} |\tau_b|\,
     \mathrm{d}\theta'  &  \mathrm{when} \quad \sigma < 0, 
   \end{array}
	\right.
\end{equation}
with
\begin{equation}
\widehat{\tau}_b = 2 \int_{\theta_{b1}}^{\theta_{b2}} |\tau_b|\, \mathrm{d}\theta.
\end{equation}
Then, the solution of (\ref{eq:mdet1}) can be easily written as
\begin{equation}
 \delta_s = 
  - \frac{I_p}{\Psi'_p}\rho_{\parallel s} 
  + \int_0^\tau \widetilde{\omega}_{s\alpha} \mathrm{d}\tau' .
  \label{eq:mdet2}
\end{equation}
In order to give a more explicit expression for (\ref{eq:mdet2}), we use periodicity in $\tau$ and write
\begin{equation}
 \widetilde{\omega}_{s\alpha} = 
  \sum_{l\neq 0} \left({\omega}_{s\alpha}\right)_l
  \mathrm{e}^{\mathrm{i}l\widehat{\omega}_b \tau}, 
\end{equation}
with $\widehat{\omega}_b := 2\pi/\widehat{\tau}_b$ and
\begin{equation}
 \left(\omega_{s\alpha}\right)_{l} = 
  (\widehat{\tau}_b/2)^{-1} \int_0^{\widehat{\tau}_b/2}
  \omega_{s\alpha}(\tau) \cos\left(l\widehat{\omega}_b \tau\right) 
  \mathrm{d}\tau.
\end{equation}
Here, the fact that $\omega_{s\alpha}(\tau)$ is even in $\tau$ has been employed. Finally,
\begin{equation}
 \delta_s =
  -\frac{I_p}{\Psi_p'}\rho_{\parallel s} 
  + 2\sum_{l>0}
  \frac{\left(\omega_{s\alpha}\right)_l}{l\,\widehat{\omega}_b} 
  \sin\left(l\, \widehat{\omega}_b \tau\right). 
  \label{eq:mdet3}
\end{equation}
Note that for axisymmetric tokamaks (\ref{eq:mdet3}) simply gives $\delta_s = -I_p\rho_{\parallel s}/\Psi'_p$.

\section{Residual zonal flow with a source term in the 
quasineutrality equation}
\label{sec:forced}

\begin{figure}[b]
\includegraphics[width=1.\linewidth]{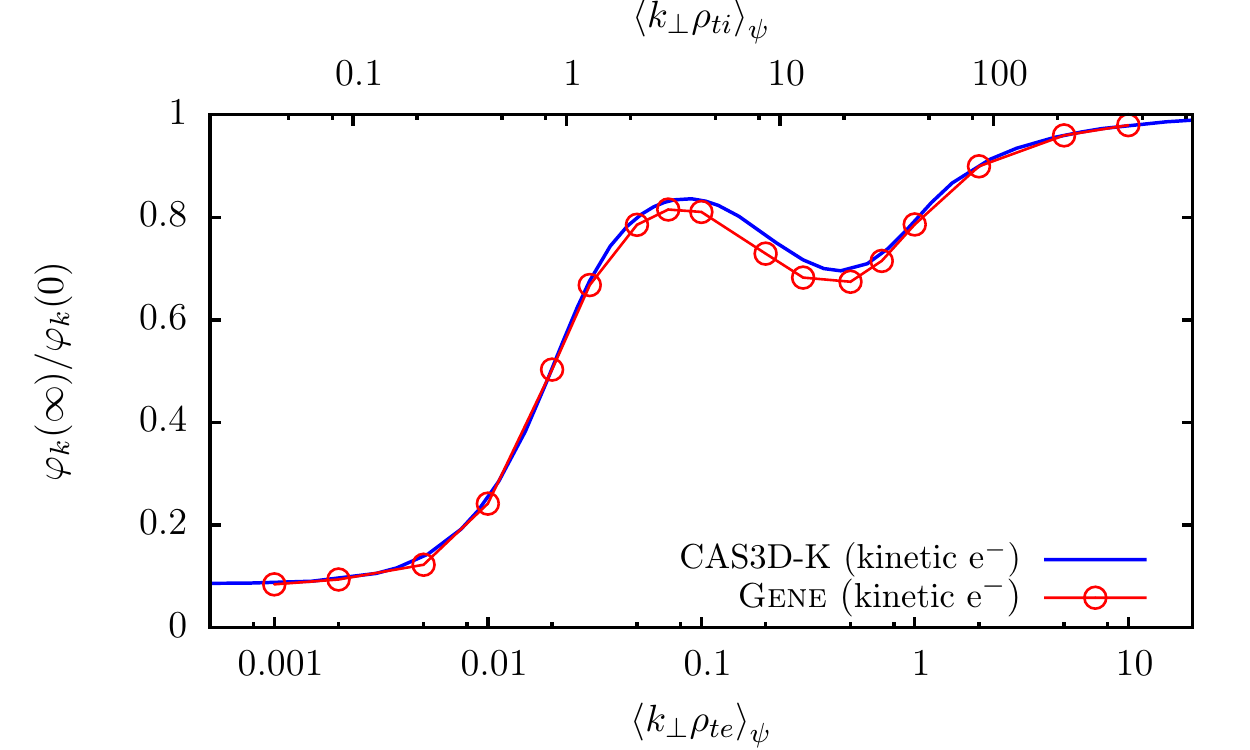}
\caption{The evaluation of equation (\ref{eq:xck}) with {\footnotesize CAS3D-K} for the tokamak of Section \ref{sec:results} at $\psi=0.25$ is shown. The corresponding simulation with {\footnotesize \gene} including a source term in the quasineutrality equation is also plotted.}
\label{fig:GENE1}
\end{figure}

\begin{figure}[t]
\includegraphics[width=1.\linewidth]{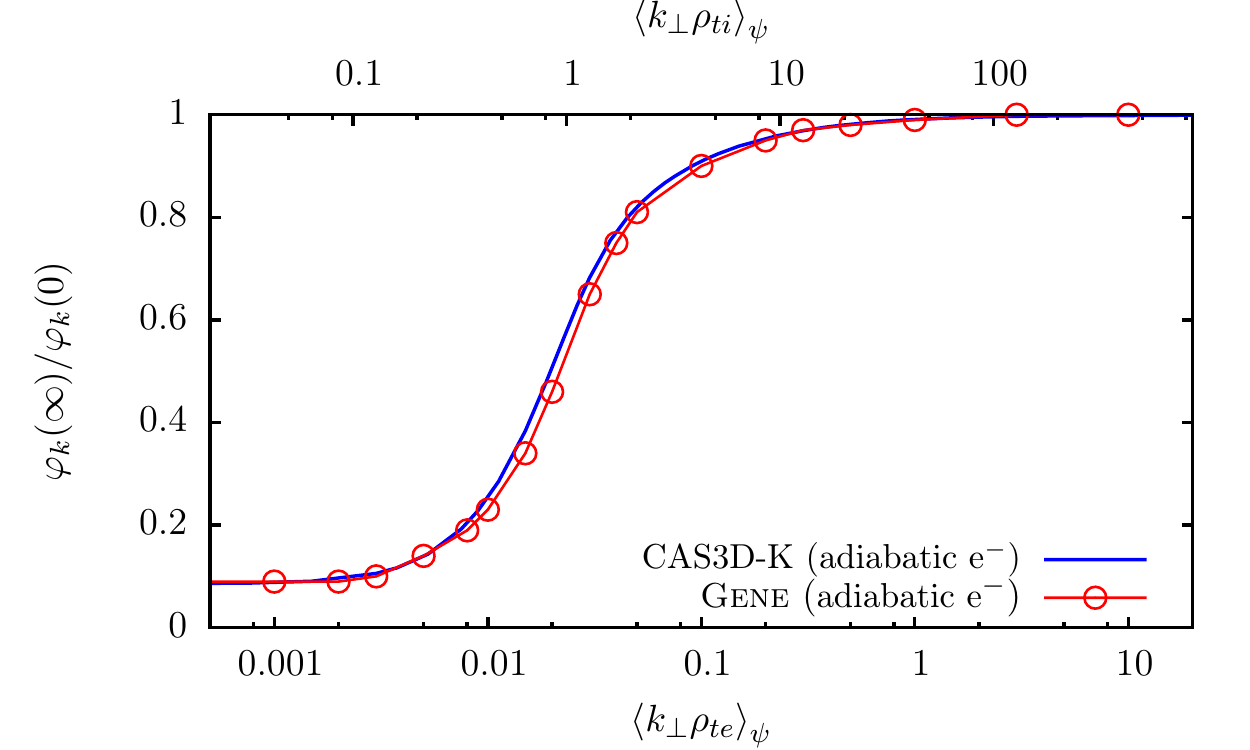}
\caption{The same calculations as in figure \ref{fig:GENE1}, but employing adiabatic electrons.}
\label{fig:GENE0}
\end{figure}

\begin{figure}[t]
\includegraphics[width=1.\linewidth]{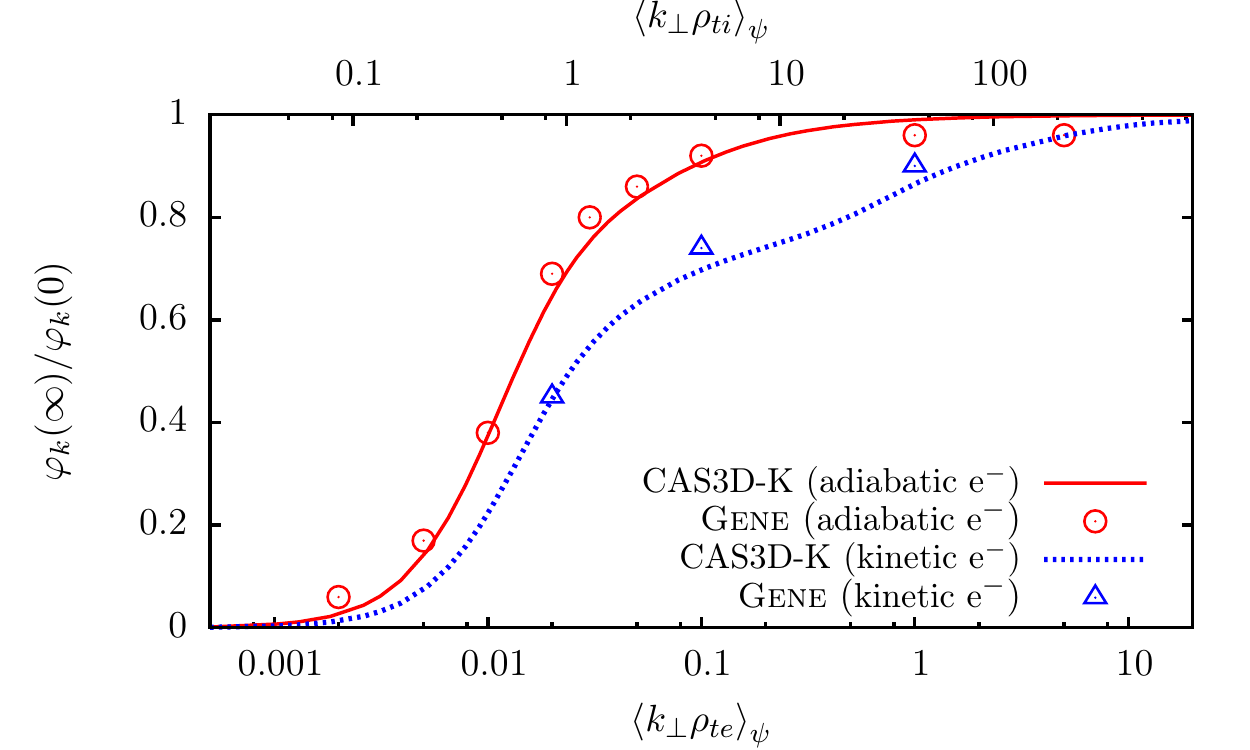}
\caption{ The results of the forced case for the standard configuration of the stellarator W7-X at $\psi=0.25$.}
\label{fig:GENE_stell_forced}
\end{figure}

Figures \ref{fig:GENE1} and \ref{fig:GENE0} are analogous to figures \ref{fig:XC2} and \ref{fig:XC1}, but this time we employ the equilibrium and parameters of Section \ref{sec:results}, and show the results obtained by both {\casdk} and \gene. The gyrokinetic simulations have been carried out by taking vanishing initial condition and adding a constant source term to the quasineutrality equation.

A formulation of the residual zonal flow problem similar to that given in \cite{Xiao2006,Xiao2007} gives, in the stellarator case,
\begin{equation}
 \frac{\varphi_k(\infty)}{\varphi_k(0)} =
  \frac{\sum_{s} \frac{Z_s^2 }{T_s}\left\{1-J_{0s}^2\right\}_s
  }{
  \sum_{s} \frac{Z_s^2}{T_s} 
  \left[\left\{1\right\}_s-\left\{
	 \mathrm{e}^{-\mathrm{i}k_\psi\delta_s} J_{0s} \, 
	 \overline{ \mathrm{e}^{\mathrm{i}k_\psi\delta_s} J_{0s}}
			   \right\}_s^{\overline{\omega_s}=0} 
  \right]
  },
  \label{eq:rlfk}
\end{equation}
if all species are kinetic. In the approximation of adiabatic electrons, one has
\begin{equation}
 \frac{\varphi_k(\infty)}{\varphi_k(0)} =
  \frac{ \sum_{s\neq e} \frac{Z_s^2 }{T_s}\left\{1-J_{0s}^2\right\}_s 
  }{
  \sum_{s\neq e} \frac{Z_s^2}{T_s} 
  \left[\left\{1\right\}_s-\left\{
	 \mathrm{e}^{-\mathrm{i}k_\psi\delta_s} J_{0s} \, 
	 \overline{ \mathrm{e}^{\mathrm{i}k_\psi\delta_s} J_{0s}}
		\right\}_s^{\overline{\omega_s}=0} 
       \right]
  }.
  \label{eq:rlfa}
\end{equation}
The evaluation of these expressions with {\casdk}, for the standard configuration of the stellarator W7-X and the parameters detailed in Section~\ref{sec:results}, is shown in \ref{fig:GENE_stell_forced}. The results for \gene~simulations are also plotted.

\section*{References}

\end{document}